\documentclass[aps,prd,11pt,a4paper,nofootinbib,oneside,superscriptaddress,eqsecnum]{revtex4-1}
\pdfoutput = 1

\usepackage{amsmath,amssymb,amsfonts,color}
\usepackage{tensor,slashed,paralist,cases,mathrsfs}
\usepackage{float,cancel,xcolor}
\usepackage{graphicx}% Include figure files
\usepackage{dcolumn}% Align table columns on decimal point
\usepackage{bm}% bold math
\usepackage{tabularx}
\usepackage{multirow}
\usepackage[colorlinks=true,urlcolor=blue,linkcolor=blue,citecolor=blue,linktocpage=true]{hyperref}

\newcommand{\beq}{\begin{eqnarray}}
\newcommand{\eeq}{\end{eqnarray}}
\newcommand{\bpmatrix}{\begin{pmatrix}}
\newcommand{\epmatrix}{\end{pmatrix}}
\newcommand{\ba}{\begin{array}}
\newcommand{\ea}{\end{array}}

\renewcommand{\eqref}[1]{Eq.~(\ref{#1})}

\newcommand{\bc}{\begin{center}}
\newcommand{\ec}{\end{center}}

\begin{document}

\vspace*{1.5em}

\title{ Freeze-out Forbidden Dark Matter in the Hidden Sector in the Mass Range from sub-GeV to TeV}

\author{Kwei-Chou Yang}
\email{kcyang@cycu.edu.tw}

\affiliation{Department of Physics and Center for High Energy Physics, Chung Yuan Christian University, 
200 Chung Pei Road, Taoyuan 32023, Taiwan}

%\date{\today $\vphantom{\bigg|_{\bigg|}^|}$}

\begin{abstract}

Kinematically forbidden channels can set the freeze-out dark matter (DM) relic abundance. These channels are described by DM annihilations into heavier states, which vanish at zero temperature limit, but occur at finite temperatures in the early Universe.
For the case that the final state of the forbidden channel is scalar mediators that couple to Standard Model (SM) matter through mixing with the SM Higgs, the signals from DM-nucleon interactions and from mediator-related missing energy or displaced vertices could be detected by direct detections and particle physics experiments, respectively. We thus present a study on the simplest secluded vector dark matter model that can exhibit this scenario in the mass range from sub-GeV to TeV.
 The dark matter resides in the hidden sector, which is in thermal equilibrium with the SM before freeze-out. During freeze-out, the depletion of its density results from its annihilation into two heavier but metastable scalars, where the coupling can be determined by having the correct relic density and constrained by the perturbative unitarity bound. However, much of the allowed parameter space is insensitive to the mixing angle between the hidden scalar and SM Higgs.
We find that a more significant mass splitting between  DM and the mediator can be allowed only in the sub-GeV region. Moreover, the mass splitting in the TeV region is required to be within the percent level. This model of the forbidden DM interacting with SM particles through the scalar portal is testable in experiments.

\end{abstract}
\maketitle
\newpage

\section{Introduction}

Despite evidence for its existence from cosmological and astrophysical measurements \cite{Adam:2015rua,Ade:2015xua},  the origin of dark matter (DM) remains unknown. 
Since DM cannot consist of any elementary particles that we have already known, its nature thus indicates the physics beyond the Standard Model (SM). In the DM search, the weakly interacting massive particles (WIMPs) serve as a popular DM candidate. Nevertheless,  the freeze-out WIMPs model is increasingly constrained due to the null measurements \cite{XENON:2018voc,PandaX-4T:2021bab, XENON:2019rxp}.  Some extension ideas based on the freeze-out scenario have recently been proposed.
One of the classes of models that can explain the relic density is the forbidden dark matter \cite{Griest:1990kh, Tulin:2012uq, Jackson:2013pjq,Jackson:2013tca, Delgado:2016umt, DAgnolo:2015ujb,DAgnolo:2020mpt, Wojcik:2021xki}. For this type of DM, during freeze-out, the process is dominated by DM annihilations into forbidden channels: $X X \to H_1 + H_2$ with $m_{H_1} +m_{H_2} >2 m_X$, which are forbidden at zero temperature. However, compared with its inverse process, although these channels are Boltzmann exponentially suppressed, they can take place at finite temperatures.

Moreover, stringent constraints on the DM from direct detections and collider experiments thus motivate us to consider the possibility that the dark matter may not couple directly to SM particles but instead interact with the SM via a mediator.
As such, for the condition that  DM is heavier than the mediator, $m_{\rm DM}>m_{\rm med}$, the DM annihilating into a metastable mediator pair can result in detectable signals in the indirect detections if it is responsible for the correct relic density (see Refs.~\cite{Yang:2019bvg, Yang:2020vxl} and references therein). On the other hand, for the condition  $m_{\rm DM}<m_{\rm med}$,   in most of the previous studies, the mediator was treated, from the kinematical consideration, as a virtual particle that mediates interactions between DM and the SM. However, in this case, the forbidden channel that the DM annihilates into the heavier mediator pair is still allowed in the early Universe. It can provide the correct relic density in some parameter space.

In this paper, we study freeze-out forbidden dark matter in a hidden sector and explore the allowed parameter space where forbidden channels mainly account for the DM freeze-out process in the early Universe.
We focus on the framework in which the dark matter ($X$) is secluded within a hidden sector.
During freeze-out, its depletion following the Boltzmann suppression is dominated by the DM annihilation into two heavier but metastable scalars $XX \to SS$, with mass $m_X < m_S$. 
In the parameter space of the present model that we will study, the following properties are satisfied before freeze-out.
\begin{enumerate}

\item The $X$ population is in thermal equilibrium with the $S$ population mainly via interactions: $XX \leftrightarrow SS$ and $X S \leftrightarrow X S$.
\item The hidden sector particles, $X$ and $S$, are in thermal equilibrium with the SM bath, where the kinetic equilibrium,
i.e., $T_X$ (hidden sector temperature) $=T$ (SM bath temperature), is maintained mainly due to elastic scatterings, $X \,  {\rm SM} \leftrightarrow X \,  {\rm SM}$ and $S \,  {\rm SM} \leftrightarrow S \,  {\rm SM}$,  and the number changing interaction, $S  \leftrightarrow   {\rm SM~SM}$ (and annihilations $SS\leftrightarrow \text{SM~SM}$,  $XX\leftrightarrow \text{SM~SM}$), while the chemical equilibrium can be kept by  $SS\leftrightarrow \text{SM~SM}$,  $XX\leftrightarrow \text{SM~SM}$ and $S\leftrightarrow \text{SM~SM}$.
\end{enumerate}

 The hidden sector could keep temperature equilibrium with the SM bath via kinetically elastic scatterings  $X \,  {\rm SM} \leftrightarrow X \,  {\rm SM}$ and $S \,  {\rm SM} \leftrightarrow S \,  {\rm SM}$ for a while. Depending on the reaction channels, the interactions, proportional to the SM density, might not be exponentially suppressed at $T<m_X, m_S$. However, a minimal coupling between the hidden sector and SM can maintain these two parts in temperature equilibrium until after freeze-out if the heating rate injected into the hidden sector by the inverse $S$ decay is larger than the cosmic cooling rate before the occurrence of the elastic decoupling. We will further discuss this point in Sec.~\ref{sec:vdm}.

 We explore the UV-complete model to provide a simple and concrete realization of this forbidden DM scenario.  In this model,  DM is an abelian gauge boson $X$ in a dark $U_X(1)$ group and gets its mass from the non-zero vacuum expectation value of the scalar $S$ in the hidden sector. In addition to the vector dark matter, this minimum extension of the Standard Model consists in adding a singlet $S$, which can mix with SM Higgs. 
The previous works in Refs.~\cite{Jackson:2013pjq, Jackson:2013tca, Delgado:2016umt} considered other models and focused on the $s$-channel resonant annihilation and/or $t$-channel annihilation of weak scale DM into the forbidden SM channels to have a correct relic density.
 On the other hand, the forbidden DM mechanism has recently been used to build models in the sub-GeV region \cite{DAgnolo:2015ujb, DAgnolo:2020mpt, Wojcik:2021xki}.  
Here we consider the forbidden model, where the DM is secluded in a hidden sector in the mass range from sub-GeV to TeV.
In our model, due to suppression of the non-relativistic DM density at $T<m_X$, the decoupling of DM annihilations to SM, in general, can occur much earlier than the freeze-out time, which is mainly determined by $XX \leftrightarrow SS$ with $m_X <m_S$.

As we will show, under the condition that the hidden sector is in thermal equilibrium with the SM before freeze-out, much of the opened parameter space where the correct relic abundance can be obtained, depending on $m_X, m_S/m_X$ and the coupling between $X$ and $S$, is insensitive to the variation of the mixing angle between the hidden scalar and SM Higgs.
The couplings of the hidden scalar to SM matter through mixing with the SM Higgs are, generally speaking, suppressed by a universal factor (more precisely, see Appendices~(\ref{app:coup-sqq})-(\ref{app:coup-sgg}) for coupling forms). For a light scalar, especially with a mass less than the $B$ meson, the LEP, LHC, and fixed target experiments can constrain the mixing angle, which may be relatively weak. For dark matter with mass from several GeV to TeV, the current direct detections and projected sensitivities can exclude partial regions of the parameter space.
Because indirect detection signals are expected to be less constraining than limits from particle physics experiments and direct detections, we will not discuss their constraints.  The parameter space of this model is still viable without violating the current experimental limits. See Sec.~\ref{sec:constraints} for detailed discussions.

The rest of this paper is organized as follows. In Sec.~\ref{thermal-freezeout}, we provide a model-independently general estimate for the freeze-out temperature parameter and $SS\to XX$ annihilation cross section from the requirement of having the correct relic density. We review aspects of the vector dark matter model and introduce the notations in Sec.~\ref{sec:vdm}, where the constraints on model parameters are discussed.  The Boltzmann equations, describing evolutions of number densities and temperature of the hidden sector particles, and related discussions are given in Sec.~\ref{sec:BoltEq}.  In Sec.~\ref{sec:constraints}, we present the experimental constraints and show the viable parameter region for this forbidden DM scenario.  We summarize in Sec.~\ref{sec:conclusions}. In Appendices, we give formulas that are relevant to this work.

\section{Estimates for thermal freezeout}\label{thermal-freezeout}

\begin{figure}[t!]
\begin{center}
\includegraphics[width=0.45\textwidth]{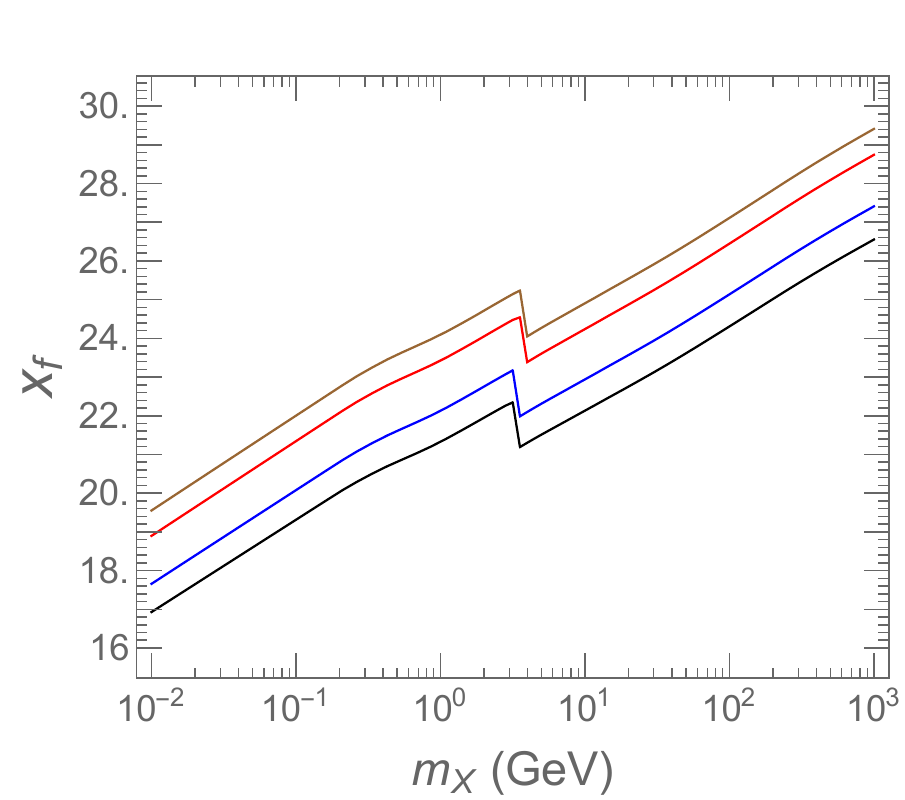}\hskip0.6cm
\includegraphics[width=0.45\textwidth]{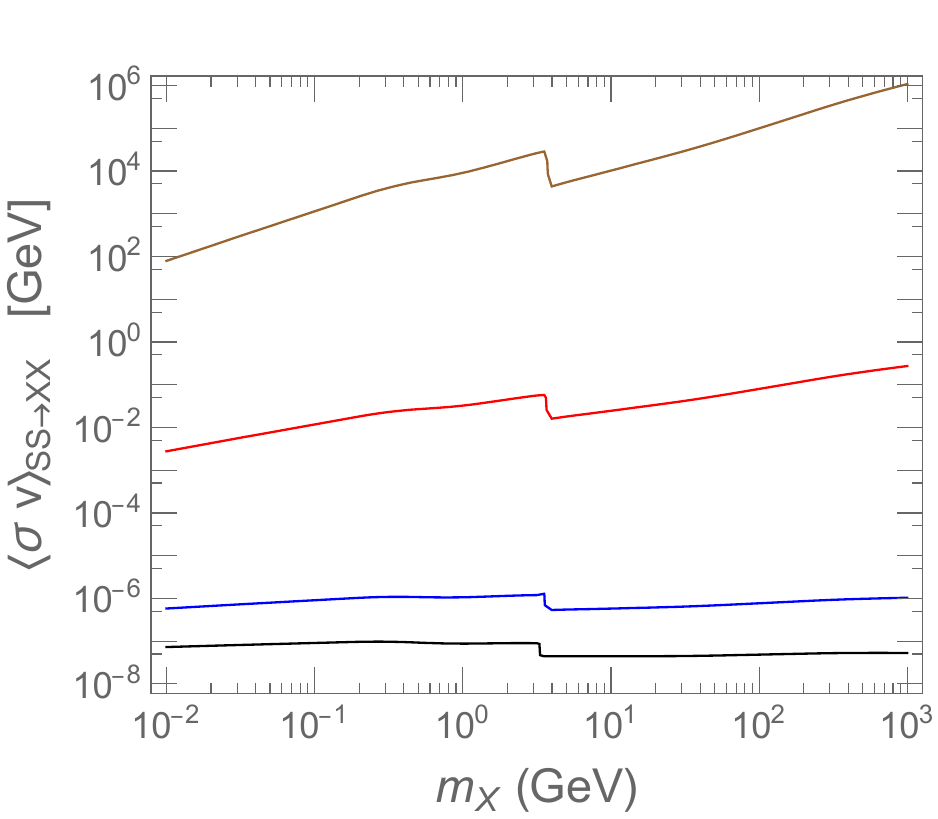}
\caption{Left panel: $x_f$ as a function of $m_X$. Right panel: $\langle \sigma v \rangle_{SS\to XX}$  that can produce the correct DM relic density as a function of $m_X$. From bottom to up, the lines (with colors black, blue, red, and brown) correspond to $r =1.01, 1.05, 1.25$, and $1.5$, respectively.
The unsmooth part of the curves denotes the QCD phase transition at temperature $T_{\rm QCD}=150$~MeV.
}
\label{fig:xf-xs-mx}
\end{center}
\end{figure}

The number density evolution of dark matter $X$ is given by the form
\begin{align}
\frac{dn_X}{dt} + 3 H n_X
=   & \bigg( \langle \sigma v \rangle_{XX\to SS}(T_S) \frac{ (n_X^{\text{eq}}(T_S))^2}{(n_S^{\text{eq}} (T_S) )^2} n_S^2
     - \langle \sigma v \rangle_{XX\to SS}(T_X) n_X^2\bigg) 
      + \dots \nonumber\\
=  & \bigg( \langle \sigma v \rangle_{SS\to XX} (T_S) \, n_S^2 - \langle \sigma v \rangle_{SS\to XX}(T_X) \, 
 \frac{  (n_S^{\text{eq}} (T_X) )^2 }{(n_X^{\text{eq}}(T_X))^2}  n_X^2 \bigg) 
+ \dots \,, 
\label{eq:evol_nx_1}
\end{align}
where, at temperature $T_i$, the detailed balance $\langle \sigma v \rangle_{SS\to XX}(T_i) \, (n_S^{\text{eq}}(T_i))^2 = \langle \sigma v \rangle_{XX\to SS}(T_i) \, (n_X^{\text{eq}}(T_i))^2$ is used, $n_{X,S}$ are the number densities for $X$ (DM) and $S$ (hidden scalar),  $n_{X,S}^{\rm eq}(T_i)$ are the corresponding equilibrium densities, and $\langle \sigma v \rangle (T_i)$ stands for the thermally averaged annihilation cross section.  In this paper, we treat that the hidden sector particles, $X$ and $S$, evolve with the same temperature, i.e., $T_X=T_S$, and in thermal equilibrium with the bath before DM freezes out.
For  $r = m_S/m_X >1$, compared with $SS \to XX$, the detailed balance condition shows that  the thermally averaged annihilation cross section for the forbidden process $XX \to SS$ is exponentially suppressed as 
 \begin{align}
\langle \sigma v \rangle_{XX\to SS}(T_X)
 &= \langle \sigma v \rangle_{SS\to XX} (T_X) \frac{(n_S^{\text{eq}}(T_X))^2}{(n_X^{\text{eq}}(T_X))^2} \nonumber \\
 & \approx \langle \sigma v \rangle_{SS\to XX}(T_X)  \frac{g_S^2}{g_X^2} r^3 
     \bigg(1- \frac{15 T_X}{4 m_X} \Big( 1- \frac{1}{r} \Big) \bigg) e^{-2(r-1)m_X/T_X} \,,
\end{align}
where $g_X$ and $g_S$ are the internal degrees of freedom of $X$ and $S$, respectively. Nevertheless, 
because the injection rate  from $S$ to $X$ is dominated by the two-body annihilation $SS\to XX$,  
the evolution process could instead satisfy $\langle \sigma v \rangle_{SS\to XX} \, n_S^2 \gtrsim 3 H n_X$ for $T \gtrsim T_f$, with $T_f = m_X/ x_f$ being the the freeze-out temperature, so that $n_X = n_X^{\rm eq}$ for $T \gtrsim T_f$. 
 To estimate the value of $x_f$, we consider that $S$ is in thermal equilibrium with the bath until the temperature below $T_f$, and $XX \leftrightarrow SS$ dominates over other number-changing interactions for $x \lesssim x_f$.  We then rewrite Eq.~(\ref{eq:evol_nx_1}), in terms of the yield $Y_X =n_X/s$, 
\begin{align}
\frac{d Y_X}{dx} \approx
   & - \frac{m_X}{x^2} \bigg( \frac{\pi}{45 G} \bigg)^{1/2} g_*^{1/2}(T_X)
 \langle \sigma v\rangle_{SS\to XX} (T_X)   \frac{(Y_S^{\text{eq}} (T_X))^2 }{ (Y_X^{\text{eq}} (T_X)^2}  
  \bigg(   Y_X^2 - \big(Y_X^{\rm eq} (T_X) \big)^2 \bigg) \,,    
  \label{eq:boltz-YX-approx}
 \end{align}
for $x \lesssim x_f$, where $Y_X^{\rm eq} =n_X^{\rm eq}/s$, $s(T) =(2\pi^2 /45) h_{\rm eff}(T) \, T^3$ is  the total entropy with $h_{\rm eff}$ the effective relativistic degrees of freedom (DoFs), and  $g_*^{1/2} \equiv \tilde{h}_{\rm eff}  /g_{\rm eff}^{1/2}$  with 
$\tilde{h}_{\rm eff}  \equiv h_{\rm eff} [1+(1/3) (d\ln h_{\rm eff} / d\ln T)]$ and $g_{\rm eff}$  the effective relativistic DoFs relevant to the total number density of the Universe.  In this paper, we compute all relevant effective DoFs adopting the QCD phase transition temperature $T_{\rm QCD}=150$~MeV (see for instance Ref.~\cite{Cerdeno:2011tf}).
Following the method given in Ref.~\cite{Gondolo:1990dk},  we define $x_f$ at which $Y_X - Y_X^{\rm eq} = \delta Y_X^{\rm eq}$ with $d\delta/dx \ll 1$ and $\delta \sim 1$ being a given number. We obtain the following approximation\footnote{$T_X=T$ before the kinetic decoupling of the DM from the bath. Meanwhile, the kinetic decoupling occurs after freeze-out. See also the discussions in Sec.~\ref{sec:BoltEq}.} from Eq.~(\ref{eq:boltz-YX-approx}),
\begin{align}
\bigg(1+ \frac{15}{8 x_f} \Big( 1- \frac{2}{r} \Big) \bigg) \bigg(1-\frac{3}{2x_f}\bigg) x_f^{1/2} e^{(2r-1)x_f}  \approx
\delta (\delta+2) \sqrt{\frac{45}{4 \pi}} \frac{M_{\rm pl} m_X r^3}{\pi^2 g_{\rm eff}^{1/2}(T_f)  } \frac{g_S^2 }{g_X}  \langle \sigma v \rangle_{SS\to XX} (T_f) \,,
\label{eq:xf}
\end{align}
where $M_{\rm pl} \equiv (8\pi G)^{-1/2}$ is the reduced Planck mass, and  $x_f$ depends only logarithmically  on $\delta$.  We will simply use $\delta=1$ in the numerical analysis.
Considering the s-wave case for the $SS \to XX$ annihilation,
we integrate Eq.~(\ref{eq:boltz-YX-approx}) from the freeze-out time $x_f$ to the present time $x_\infty$,
\begin{align}
\int^{Y_X^{\infty}}_{Y_X^f}  \frac{d Y_X}{Y_X^2} \approx 
- m_X \bigg( \frac{\pi}{45 G} \bigg)^{1/2} \overline{g_*}^{1/2} \langle \sigma v\rangle_{SS\to XX}(T_f)   
\int^{x_\infty}_{x_f} 
 \frac{(Y_S^{\text{eq}} (T_X))^2 }{ (Y_X^{\text{eq}} (T_X)^2}   \frac{dx}{x^2} \,,
\end{align}
where,  during the integral range, $Y_X^{\rm eq} /Y_X \ll 1$ is used, $T_X$ is approximated as $T$, and $\overline{g_*}^{1/2}$ is a typical value of $g_*^{1/2}$, which will be simply taken $\overline{g_*}^{1/2} = g_*^{1/2}(T_f)$ in the analysis.
Further using the properties $x_\infty \gg x_f$ and $Y_X^\infty \ll Y_X^{f}$, we have 
\begin{align}
\frac{1}{m_X Y_X^{\infty}} 
\approx  
&\frac{4\pi}{\sqrt{90}}M_{\rm pl} g_*^{1/2}(T_f)
 \langle \sigma v\rangle_{SS\to XX} (T_f)   
 \frac{g_S^2 }{g_X^2}  r^3 
  %\times 
  \bigg\{  \frac{e^{-2(r-1)x_f}}{x_f}  +  2(r-1)   {\rm E_i}\big(\! -2(r-1)x_f \big)
\nonumber\\
&   -\frac{15}{8}\Big( 1-\frac{1}{r} \Big) 
 \bigg[ \frac{1-2(r-1)x_f}{x_f^2} e^{-2(r-1)x_f} -  \big( 2(r-1) \big)^2{\rm E_i}\big(\! -2(r-1)x_f \big)
    \bigg] \bigg\} \,,
 \label{eq:xs}
\end{align}
where the exponential integral ${\rm Ei}(z)$ is defined as ${\rm Ei}(z)= - \int_{-z}^{\infty} (e^{-t}/t) dt$.
 The  post-freeze-out value of the DM yield $ Y_X^{\infty}$ is related to the DM relic abundance \cite{pdg2022},
\begin{align}
\Omega_X =  \frac{m_X Y_X^\infty  s_0}{\rho_c} \simeq 0.1198/h^2 \,,
\end{align} 
where $\rho_c = 1.0537\times 10^{-5} h^2 ({\rm GeV}/c^2) {\rm cm}^{-3}$ is the present critical density, $h\simeq 0.674$ is  the scale factor for Hubble expansion rate, and
$s_0=2891~\text{cm}^{-3}$ is the present-day entropy. For the results shown in Eqs.~(\ref{eq:xf}) and (\ref{eq:xs}), we have used the approximation,
\begin{align}
n_i^{\rm eq}(T_f) = g_i \frac{m_i^2 T_f}{2 \pi^2} K_2 (m_i/T_f) \,,
\end{align}
with the modified Bessel function of the second kind
\begin{equation}
K_2( m_i/T_f ) \simeq \sqrt{ \frac{\pi}{2(m_i/m_X) x_f} } \bigg(1+ \frac{15}{8 (m_i/m_X) x_f} \bigg) e^{-(m_i/m_X) x_f} \,,
\end{equation}
and $i\equiv X, S$.
Solving the two Eqs.~(\ref{eq:xf}) and (\ref{eq:xs}) numerically, we show in Fig.~\ref{fig:xf-xs-mx} the results of  $x_f$ and $\langle \sigma v \rangle_{SS\to XX}$, respectively, as functions of $m_X$, with respect to different values of $r=1.01, 1.05, 1.25$, and $1.5$. The value of $x_f$ is in the range $17\sim 29$ corresponding to $m_X \in$ [10~MeV, 1~TeV].  Compared with the s-wave DM annihilation cross in the WIMP case, which is of order $10^{-9}$ GeV, the value of $\langle \sigma v \rangle_{SS\to XX}$ that, especially depending on the value of $r$, can produce the correct DM relic density, is in the range of $10^{-7}\sim 10^{6}$~GeV$^{-2}$.

\section{The vector dark matter model}\label{sec:vdm}

In the simplest renormalizable vector dark matter model,  the vector dark matter, $X$, is the gauge boson of a dark $U(1)_X$ (see, e.g., Refs.~\cite{Baek:2012se, Yang:2019bvg, Yang:2020vxl}).
The dark kinetic terms and scalar potentials of the  Lagrangian are given by
\begin{align}
 {\cal L}_\text{hidden} = & -\frac{1}{4} (\partial_\mu X_\nu -\partial_\nu X_\mu)^2 + (D_\mu\Phi_S)^\dagger (D^\mu \Phi_S) 
    - \mu_{H}^2 |\Phi_H|^2 - \mu_{S}^2 |\Phi_S|^2 \nonumber \\  
 &- \frac{\lambda_H}{2} (\Phi_H^\dagger \Phi_H)^2 
 - \frac{\lambda_S}{2} (\Phi_S^\dagger \Phi_S)^2 
 - \lambda_{HS} (\Phi_H^\dagger \Phi_H) (\Phi_S^\dagger \Phi_S)  \;,
\label{eq:lagrangian}
\end{align}
where  $ D_\mu \Phi_S \equiv (\partial_\mu + i g_{\rm dm} Q_{\Phi_S} X_\mu )\Phi_S$,  the SM Higgs doublet $\Phi_H= (H^+, H^0)^{\rm T}$, and the complex scalar $\Phi_S$ is charged under a $U_X(1)$ gauge symmetry with charge $Q_{\Phi_S}$.   For simplicity, we will use $Q_{\Phi_S}=1$.
After spontaneous symmetry breaking, the $Z_2$ symmetry, $X_\mu \to -X_\mu$ and $\Phi_S \to \Phi_S^*$, persists, making $X_\mu$ stable. 
 The two neutral Higgs fields are parametrized as
\begin{equation}
 H^0=\frac{1}{\sqrt{2}} (v_H + \phi_h + i \sigma_h), \quad
\Phi_S=\frac{1}{\sqrt{2}} (v_S + \phi_s + i \sigma_s).
\label{eq:vev}
\end{equation}
Thus $\sigma_s$ becomes the longitudinal component of $X$, and vector DM has the mass $m_X=g_{\rm dm} Q_{\Phi_S} v_S$.
The mass eigenstates of physical Higgses $(h, S)$ are given by
\begin{align}
h & =\ \ \cos\alpha \, \phi_h  +  \sin\alpha \, \phi_s \,, \\
S &=- \sin\alpha \, \phi_h  + \cos\alpha \, \phi_s \,, 
\end{align}
and the mass term in the Lagrangian is then rewritten as $ -1/2\, (\phi_h, \phi_s)\,  M_{\rm Higgs}^2 \,  (\phi_h, \phi_s)^\dagger $ with
\begin{equation}
M_{\rm Higgs}^2 =
\left(
\begin{array}{cc}
 m_h^2 \cos^2\alpha  + m_S^2 \sin^2\alpha  & (m_h^2 -m_S^2 )\sin\alpha \cos\alpha \\ 
(m_h^2 -m_S^2 )\sin\alpha \cos\alpha & m_S^2 \cos^2\alpha + m_h^2 \sin^2\alpha
\end{array}
\right) \,.
\label{eq:mass_matrix}
\end{equation}
The particle $h$ is identified as the SM-like Higgs with mass $m_h=125.18$~GeV, and the vacuum expectation $v_H$ is fixed to be $v_H=246$~GeV.

\begin{figure}[t!]
\begin{center}
\includegraphics[width=0.46\textwidth]{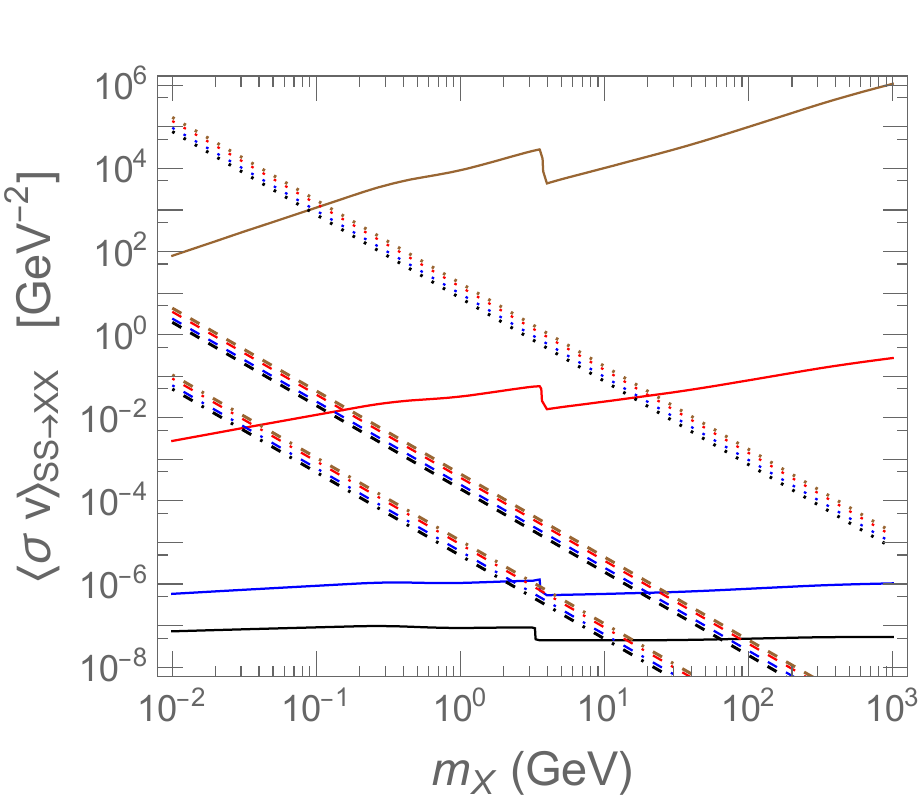}\hskip0.6cm
\includegraphics[width=0.456\textwidth]{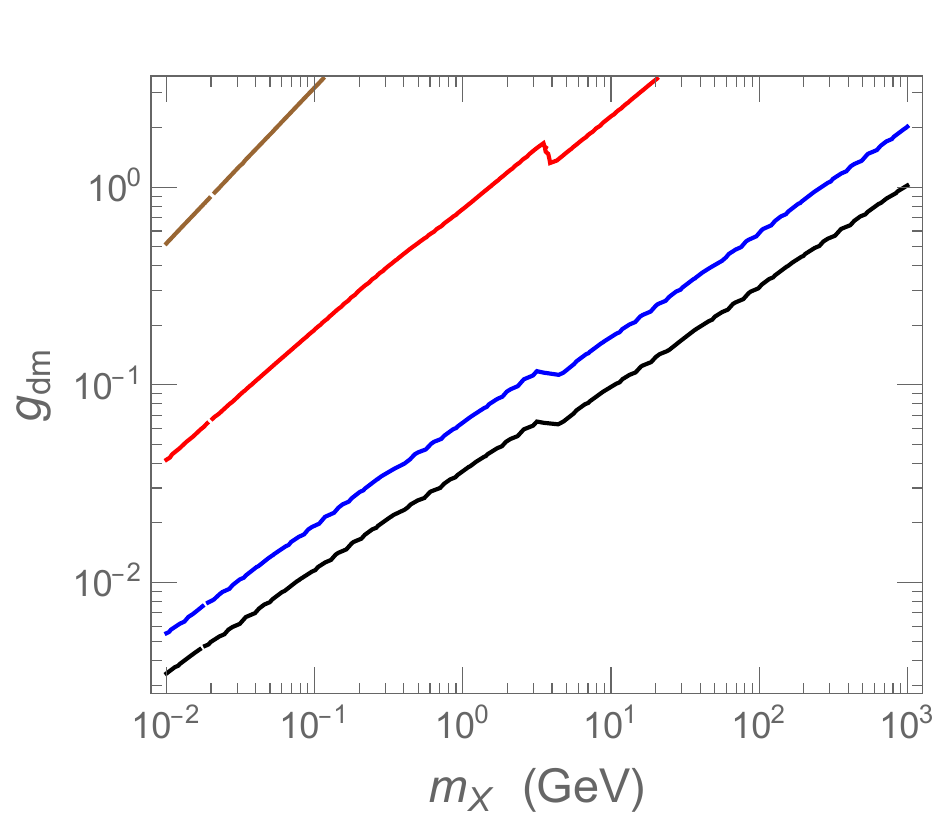}
\caption{Left panel: thermal cross section $\langle \sigma v \rangle_{SS\to XX}$ at $T=T_f$ as a function of $m_X$. Dotted, dashed, and dotdashed lines are thermal cross sections in the vector dark matter model using the inputs $g_{\rm dm} = \sqrt{4\pi}, 0.252$ and $0.1$, respectively, in comparison with the results that can produce correct relic density (solid lines in the horizontal direction)  as shown in the right panel of Fig.~\ref{fig:xf-xs-mx}. Right panel:  coupling constant $g_{\rm dm}$ as a function of $m_X$, that can produce the correct relic density as discussed in Sec.~\ref{thermal-freezeout}.
Again, from bottom to up, the lines (with black, blue, red, and brown) correspond to $r (=m_S/m_X)=1.01, 1.05, 1.25, 1.5$, respectively.}
\label{fig:model-xs-mx-gdm}
\end{center}
\end{figure}

Using the fixed mass ratio $r (=m_S/m_X)$,  we show the thermal freeze-out cross section $\langle \sigma v \rangle_{SS\to XX}(x_f) $  in left panel of Fig.~\ref{fig:model-xs-mx-gdm}, with respect to $g_{\rm dm}= \sqrt{4\pi}$ (the perturbative unitarity bound), $0.252$, and $0.1$, respectively, in comparison with the results that can produce the correct relic density. In the right panel of Fig.~\ref{fig:model-xs-mx-gdm}, adopting the freeze-out $SS\to XX$ annihilation cross section which can produce the correct relic density, estimated from Eqs.~(\ref{eq:xf}) and (\ref{eq:xs}), we depict the allowed coupling constant $g_{\rm dm}$ as a function of $m_X$. We find that a larger mass-splitting ratio between  DM and the hidden scalar can be allowed only in the sub-GeV region. Moreover, the mass splitting for $m_{X,S} \gtrsim O({\rm TeV})$ is significantly below the percent level, by imposing the requirement of the perturbative unitarity bound $g_{\rm dm} < \sqrt{4\pi}$.

Since $\alpha$ is relevant to the coupling strengths of the hidden sector to SM particles, we can thus restrict the value of the range in which these two parts are well in kinetic equilibrium with each other before freeze-out. 
This equilibrium is mainly because the kinetic energy injection rate to the hidden sector either via elastic scatterings, $S~{\rm SM} \to S~{\rm SM}$ and  $X~{\rm SM} \to X~{\rm SM}$,  or via the inverse decay, ${\rm SM~SM} \to S$, is larger than the Hubble cooling rate.

\begin{figure}[t!]
\begin{center}
\includegraphics[width=0.55\textwidth]{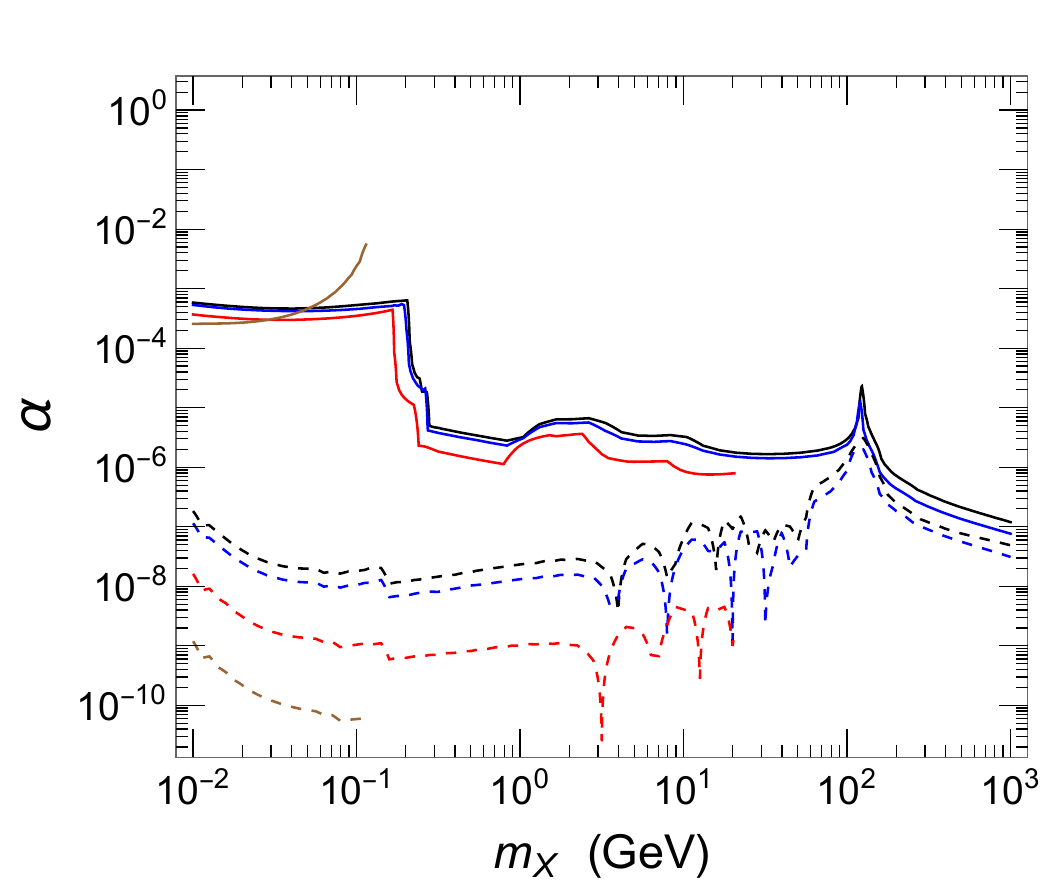}
\caption{ (a) The solid lines show the lower bound of $\alpha$ about keeping the hidden sector in temperature equilibrium with the thermal bath before freeze-out, assuming that the hidden sector particles follow the equilibrium density distributions; see the discussions in Sec.~\ref{sec:vdm}.
(b) The dashed lines show the minimum values of $\alpha$ above that $Y_{X,S}$ grow by up to $Y_{X,S}^{\rm eq}(T)$ at $T$ with $T\gtrsim m_X$; see Eq.~(\ref{eq:freezein}) and its related discussions.
From up to bottom at $m_X=0.02$~GeV the lines (with  colors black, blue, red, and brown) correspond to $r (=m_S/m_X)=1.01, 1.05, 1.25, 1.5$, respectively.
}
\label{fig:alp_mx}
\end{center}
\end{figure}

In our setup, we adopt the lower bound of the $\alpha$ range to be the value at which the heating rate transferred to the hidden sector via the inverse $S$ decay  is equal to the cooling rate due to the cosmic expansion, $\langle\Gamma_S\rangle \delta_\Gamma n_S^{\rm eq} T = [(2 -\delta_H^S) n_S + (2 -\delta_H^X) n_X ] H T_X$, at $T= T_{\rm el}$,  but it becomes $\langle\Gamma_S\rangle \delta_\Gamma n_S^{\rm eq} T >   [(2 -\delta_H^S) n_S + (2 -\delta_H^X) n_X ] H T_X$ for $T < T_{\rm el}$, where $\langle\Gamma_S\rangle$ is the thermally averaged decay width of $S$, and $T_{\rm el}$ is the elastic decoupling temperature of the hidden sector from the SM bath determined by  interactions $X \,  {\rm SM} \leftrightarrow X \,  {\rm SM}$ and $S \,  {\rm SM} \leftrightarrow S \,  {\rm SM}$. See also the discussion in Sec.~\ref{sec:BoltEq}.  The result is depicted in Fig.~\ref{fig:alp_mx}. Here we have used $T_X=T, n_X=n_X^{\rm eq}$ and $n_S=n_S^{\rm eq}$  before freeze-out (see Eq.~(\ref{eq:freezein}) for further discussions). The definitions of $\delta_\Gamma, \delta_H^S$ and  $\delta_H^X$ can be found in Eqs.~(\ref{app:deltag}) and (\ref{eq:deltah}).

In addition, in Fig.~\ref{fig:alp_mx}, we also show the minimum values of $\alpha$ above that $Y_{X,S}$ can grow by up to $Y_{X,S}^{\rm eq}(T)$ at $T$ with $T\gtrsim m_X$. The details will be given in Sec.~\ref{sec:BoltEq} (see Eq.~(\ref{eq:freezein}) and its related discussions).

Before closing this section, two remarks are in order as follows. First, as we will show in the next section, even using the lower bound value of $\alpha$ for $r\lesssim 1.05$, the number densities of $X$ and $S$ can follow quite well their equilibrium values with zero chemical potential before freeze-out. It means that the interactions, $SS\leftrightarrow \text{SM~SM}$,  $XX\leftrightarrow \text{SM~SM}$ and $S\leftrightarrow \text{SM~SM}$, can maintain the chemical equilibrium between the hidden sector and SM in the parameter region above the lower bound of $\alpha$.
See also Figs.~\ref{fig:relic-boltz1-1} and \ref{fig:relrate} for reference. Second, as seen in the right panel of Fig.~\ref{fig:model-xs-mx-gdm}, $g_{\rm dm}$ increases with $r$
for a fixed value of $m_X$. We will show in the next section that for a smaller $r\lesssim 1.05$, the value of  $g_{\rm dm}$ as a function of $m_X$ is almost independent of $\alpha$. However, we find that, for a larger $r$, e.g., $r=1.25$, using the lower bound value of $\alpha$ may result in the hidden sector being out of kinetic equilibrium with the SM bath before freeze-out. Therefore, through the interaction $S\leftrightarrow \text{SM~SM}$, we need to enlarge the $\alpha$ value to maintain the kinetic equilibrium between the two parts well until after freeze-out to account for the correct relic density.

\section{Boltzmann equations}\label{sec:BoltEq}

More precise estimates for model parameters can be obtained by solving Boltzmann equations.
Since the evolutions of the number densities of the hidden sector particles depend on their temperature, we also need to consider their temperature Boltzmann equation.

The following coupled Boltzmann equations describe the evolutions of number densities of $X$ and $S$,
\begin{align}
\frac{dn_X}{dt} + 3 H n_X=  
& \bigg( \langle \sigma v \rangle_{SS\to XX} (T_S) \, n_S^2 - \langle \sigma v \rangle_{SS\to XX}(T_X) \, 
 \frac{  (n_S^{\text{eq}} (T_X) )^2 }{(n_X^{\text{eq}}(T_X))^2}  n_X^2 \bigg) 
    + \{3 \leftrightarrow 2 \}_X    \nonumber\\
%    -  \frac{1}{3} \langle \sigma v^2 \rangle_{XXX\to XS} \bigg( n_X^3 -n_X n_S \frac{(n_X^{\text{eq}})^2}{n_S^{\text{eq}}} \bigg)   \nonumber\\
% & -  \langle \sigma v^2 \rangle_{XXS\to SS} \bigg( n_X^2 n_S -(n_X^{\text{eq}})^2 \frac{n_S^2}{(n_S^{\text{eq}})} \bigg)
%   + \frac{1}{3} \langle \sigma v^2 \rangle_{SSS\to XX} \bigg( n_S^3 -(n_S^{\text{eq}})^3 \frac{n_X^2}{(n_X^{\text{eq}})^2} \bigg) 
%   \nonumber\\
  &  -   \bigg( \langle \sigma v \rangle_{XX\to  \sum_{ij} {\rm SM}_i\, {\rm SM}_j} (T_X) \,  n_X^2
     -   \langle \sigma v \rangle_{XX\to  \sum_{ij}{\rm SM}_i\, {\rm SM}_j} (T)  \, (n_X^{\text{eq}} (T) )^2   \bigg) 
   \,, \label{eq:boltz-1} 
\end{align}
%\\
\begin{align}
\frac{dn_S}{dt} + 3 H n_S = & -  \Gamma_{S}
        \bigg( \frac{K_1(x_S\cdot m_S/m_X)}{K_2(x_S\cdot m_S/m_X)} n_S -  \frac{K_1(x\cdot m_S/m_X) }{K_2 (x\cdot m_S/m_X)}  n_S^{\text{eq}} (T)  \bigg)  \nonumber\\
  &  -   \bigg( \langle \sigma v \rangle_{SS\to  \sum_{ij} {\rm SM}_i\, {\rm SM}_j} (T_S)\,  n_S^2
     -   \langle \sigma v \rangle_{SS\to  \sum_{ij}{\rm SM}_i\, {\rm SM}_j} (T) (n_S^{\text{eq}} (T) )^2   \bigg) \nonumber\\
 & -  \bigg( \langle \sigma v \rangle_{SS\to XX} (T_S) \, n_S^2 
                 - \langle \sigma v \rangle_{SS\to XX}(T_X) \,    \frac{ (n_S^{\text{eq}} (T_X) )^2 }{(n_X^{\text{eq}} (T_X))^2} n_X^2 \bigg)
       + \{3 \leftrightarrow 2 \}_S   \,, \label{eq:boltz-2} 
%      + \frac{1}{6} \langle \sigma v^2 \rangle_{XXX\to XS} \bigg( n_X^3 -n_X n_S \frac{(n_X^{\text{eq}})^2}{n_S^{\text{eq}}} \bigg)  \nonumber\\
% & + \frac{1}{2} \langle \sigma v^2 \rangle_{XXS\to SS} \bigg( n_X^2 n_S -(n_X^{\text{eq}})^2 \frac{n_S^2}{(n_S^{\text{eq}})} \bigg)
%   - \frac{1}{2} \langle \sigma v^2 \rangle_{XSS\to XS} \bigg( n_X n_S^2 - n_X n_S n_S^{\text{eq}}   \bigg) 
% \nonumber\\
% &  - \frac{1}{2} \langle \sigma v^2 \rangle_{SSS\to XX} \bigg( n_S^3 -(n_S^{\text{eq}})^3 \frac{n_X^2}{(n_X^{\text{eq}})^2} \bigg)
%  - \frac{1}{6} \langle \sigma v^2 \rangle_{SSS\to SS} \bigg( n_S^3 -n_S^2 n_S^{\text{eq}}   \bigg)  \,, \label{eq:boltz-2}
\end{align}
where $\{3 \leftrightarrow 2 \}_{X}$ and $\{3 \leftrightarrow 2 \}_{S}$  denote terms involving $3 \leftrightarrow 2$ number changing interactions.
The detailed expressions for the $2 \to 2 $ annihilation cross sections and $3 \leftrightarrow 2 $ interaction terms appearing in the above equations are collected in Appendices~\ref{app:2to2} and \ref{app:boltz1}, respectively. 
We define the temperature of the hidden sector particles as 
\begin{align}
T_i =\frac{g_i}{n_i (T_i)} \int \frac{d^3 p_i}{(2\pi)^3} \frac{{\bf p}_i^2}{3 E_i}  f_i (T_i) \,
\label{eq:temp-def}
\end{align} 
 (with $i\equiv X$ or $S$).
 This is a good approximation for the present study that the reactions are mostly contributed by the phase space region $f_i \ll1$, and therefore 
\begin{equation}
f_{i} =  e^{-(E_{i}-\mu_i)/T_i} (1 + f_i) \simeq e^{-(E_{i}-\mu_i)/T_i}  \,.
\label{eq:dist-approx}
\end{equation}
See also the discussion in Appendix~\ref{app:col-T}.

The temperature evolutions derived from the Boltzmann moment equations for $T_X$ and $T_S$ are given in the form,
\begin{align}
    n_X \bigg( \frac{d T_X}{dt} + (2-\delta_H^X) H T_X  \bigg) 
      = &      - \left( \frac{d n_X }{dt} +3 H n_X  \right) T_X 
          + g_X \int \frac{d^3p_X}{(2\pi)^3}  C \Big[ f_X\cdot \frac{{\bf p}_X^2}{3 E_X} \Big] 
\,,  \label{eq:boltz-t-1}
\\
%\end{align}
%
%\begin{align}
  n_S   \bigg( \frac{d T_S}{dt} + (2-\delta_H^S) H T_S  \bigg)  
      =&  - \left( \frac{d n_S }{dt} +3 H n_S \right) T_S 
          + g_S \int   \frac{d^3p_S}{(2\pi)^3}   C \Big[f_S\cdot \frac{{\bf p}_S^2}{3 E_S} \Big] 
  \,,  \  
  \label{eq:boltz-t-2}
\end{align}
where 
\begin{align}
\delta_H^i 
& \equiv 1- \frac{g_i}{n_i (T_i) \, T_i} \int \frac{d^3 p_i}{(2\pi)^3} \frac{{\bf p}_i^2 m_i^2}{3 E_i^3}  f_i (T_i) \nonumber\\
& \to 1- \frac{\int_1^\infty   d\alpha\, e^{-\alpha x} (\alpha^2 -1)^{3/2} / \alpha^2}{\int_1^\infty d\alpha\, e^{-\alpha x} (\alpha^2 -1)^{3/2}}  
\qquad \text{for  $n_i(T_i) \to n_i^{\rm eq}(T_i)$  }\,,
\label{eq:deltah}
\end{align} 
$f_{i}$ and $g_i$  are the distribution and internal degrees of freedom of the hidden sector particles in the phase space, respectively, and
$g_i \int    C \big[f_i \cdot \frac{{\bf p}_i^2}{3 E_i} \big]     d^3p_i/(2\pi)^3  $ 
is the collision term, of which the detailed forms are given in Appendix~\ref{app:col-T}.
Here $\delta_H^{X(S)}$ is approximately equal to 0.9 for $T=m_{X(S)}$, and approaches to 0 or 1 for $T\ll m_{X(S)}$ or $T\gg m_{X(S)}$. 
In this work, we tend to consider that the $X$ and $S$ particles evolve with the same temperature, i.e., $T_X=T_S$, and their kinetic decoupling occurs after the DM freeze-out. 
We treat the hidden sector as a whole to determine its temperature evolution which is obtained by taking the sum of Eqs.~(\ref{eq:boltz-t-1}) and (\ref{eq:boltz-t-2}). The resulting temperature evolution is numerically stable, given by
\begin{align}
      \frac{d T_X}{dt}   
      +  \frac{1}{n_X  + n_S  }
     \Big( & (2-\delta^X_H) n_X +   (2-\delta^S_H) n_S  \Big)   H  T_X  \nonumber  \\
   =   \frac{1}{n_X  + n_S  }
     \bigg[ &- \left( \frac{d n_X }{dt} +3 H n_X  \right) T_X
          + g_X \int  \frac{d^3p_X}{(2\pi)^3}  \, C \Big[f_X\cdot \frac{{\bf p}_X^2}{3 E_X} \Big]  \nonumber\\
              & - \left( \frac{d n_S }{dt} +3 H n_S  \right) T_X 
          + g_S \int  \frac{d^3p_S}{(2\pi)^3}  \, C \Big[f_S\cdot \frac{{\bf p}_S^2}{3 E_S} \Big] 
       \bigg]
     \,.
       \label{eq:boltz-t-hidden}
\end{align}
In the study, we instead use dimensionless quantities,
\begin{align}
&Y_X \equiv \frac{n_X}{s} \,, \qquad  Y_S \equiv \frac{n_S}{s}\,,    \qquad {\rm (yields)}  \nonumber\\
& y= \frac{T_X}{T}\,,     \qquad {\text{(temperature ratio)}}
\label{eq:dim-less-variables}
\end{align}
as functions of the dimensionless variable $x \equiv m_X/T$,
where $s$ is the total entropy density of the Universe. The formulas for the Boltzmann equations of $Y_X, Y_S$, and $y$ are given in Appendix~\ref{app:boltz}.

\begin{figure}[t!]
\begin{center}
\vskip-0.35cm
\includegraphics[width=0.457\textwidth]{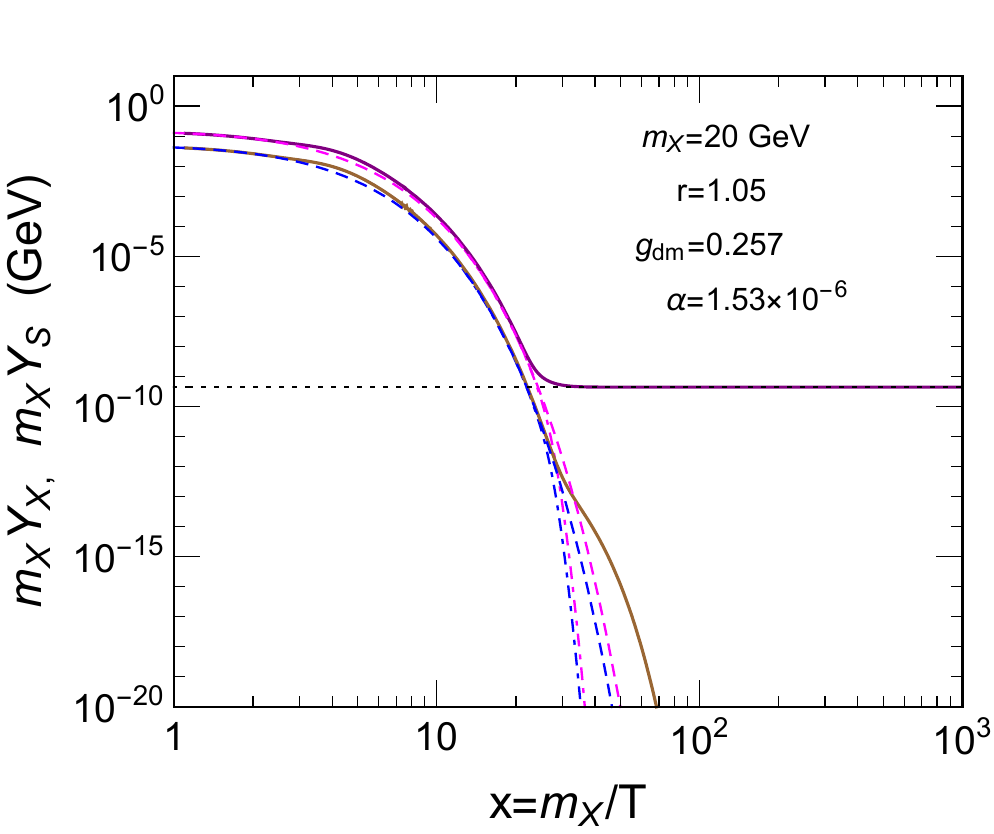}\hskip0.5cm
\includegraphics[width=0.45\textwidth]{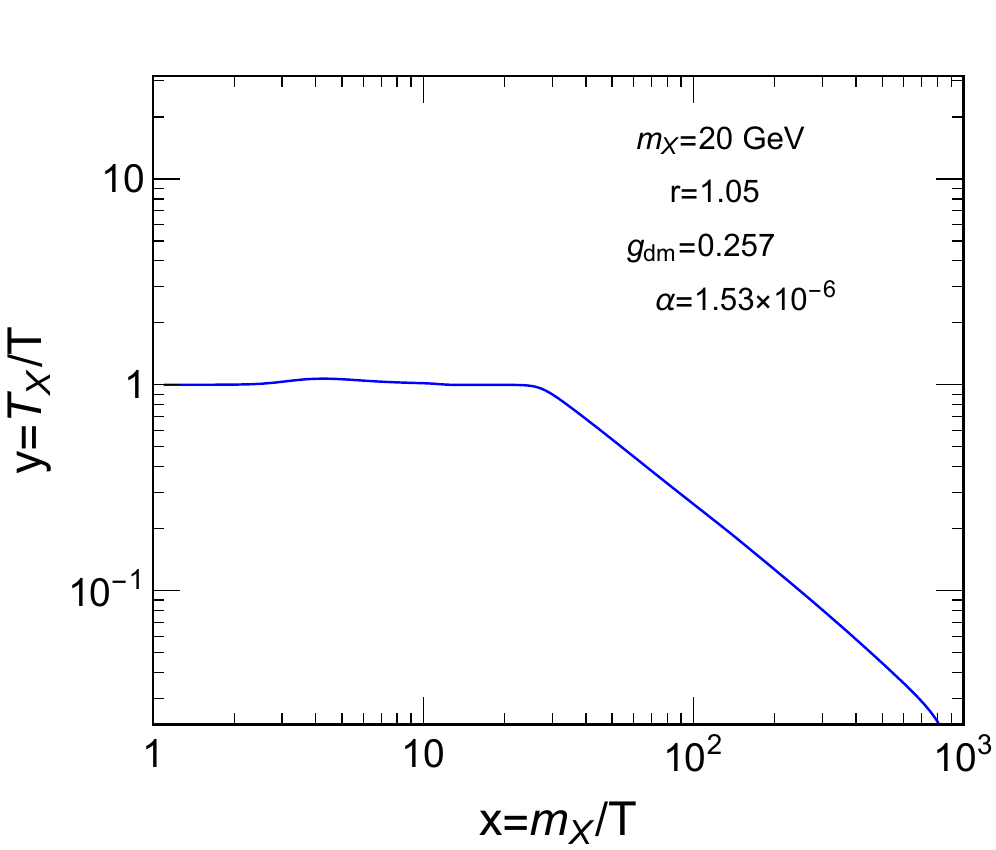}\\
\includegraphics[width=0.457\textwidth]{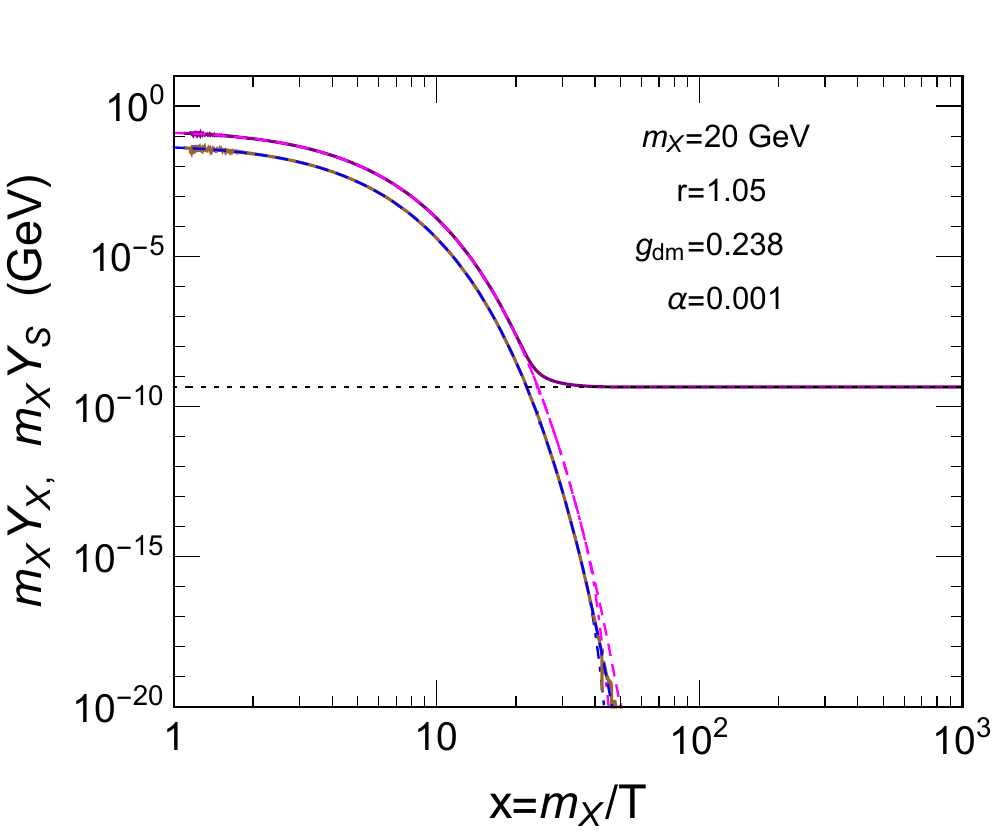}\hskip0.5cm
\includegraphics[width=0.45\textwidth]{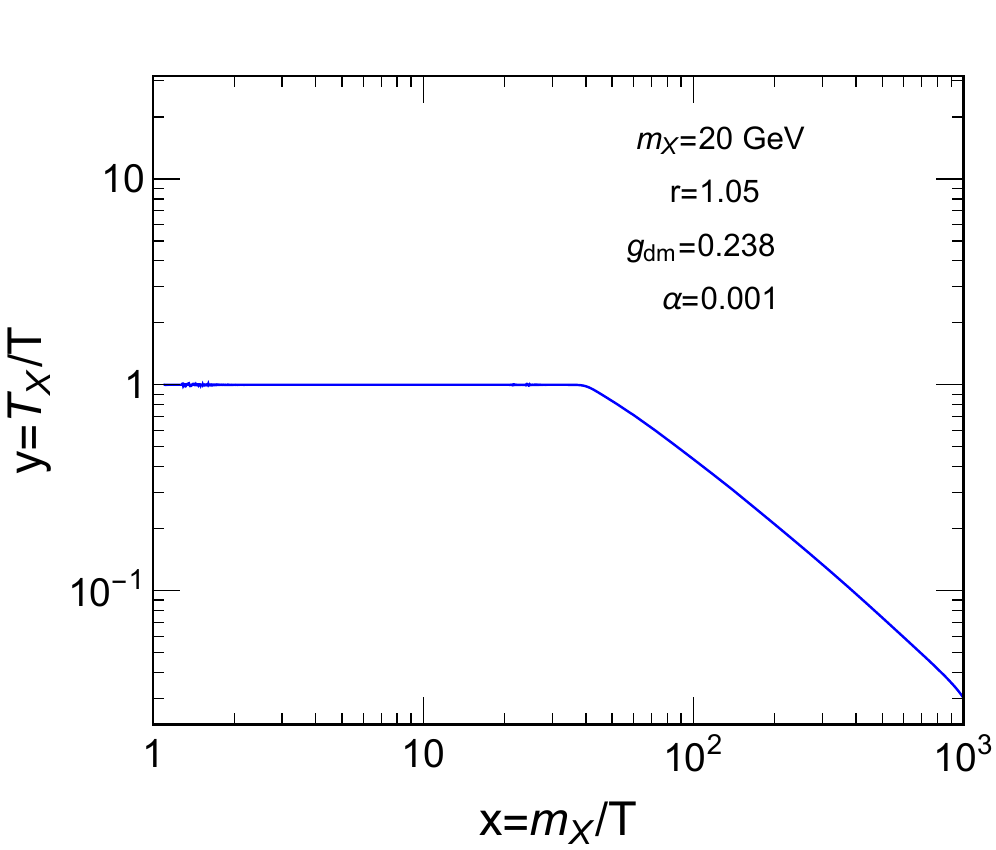}\\
\vskip-0.35cm
\caption{Right panel:  $y=T_X/T (=T_S/T)$ vs. $x (=m_X/T)$, where $T$ is temperature of the bath.  Left panel:
$m_X Y_X$ (purple solid line) and $m_X Y_S$ (brown solid line) as functions of $x$. The magenta and blue dashed lines (or dotdashed lines) depict the corresponding ones following Boltzmann suppression with $T_{X}=T$ (or with their actual temperature).  Here, the DM freeze-out yield can lead to the correct relic abundance denoted by the horizontal dotted line.
}
\label{fig:relic-boltz1-1}
\end{center}
\end{figure}

We focus on the scenario where the hidden sector particles are in temperature equilibrium with the SM bath before freeze-out.
 In Fig.~\ref{fig:relic-boltz1-1}, we show the thermodynamic evolutions of the hidden sector species for the case of $m_X=20$~GeV and $r (=m_S/m_X) =1.05$. 
We first use $\alpha= 1.53\times 10^{-6}$ as an input parameter, which, as shown in Fig.~\ref{fig:alp_mx} and discussed in the previous section, is the lower bound corresponding to the chosen values of $m_X$ and $r$.  We find that $g_{\rm dm}=0.257$ can produce the correct relic density $\Omega_X \simeq 0.1198/h^2$.
 In Fig.~\ref{fig:relic-boltz1-1}, both yields for $X$ and $S$ are scaled by a factor of $m_X$, while the horizontal dashed line denotes the value of the correct relic abundance, which is equal to $m_X Y_X^\infty=4.38\times 10^{-10}$~GeV and is independent of the value of $m_X$. 

We also show evolutions in Fig.~\ref{fig:relic-boltz1-1}  by using a much larger value of $\alpha=0.001$ and find that $g_X=0.238$ can account for the correct relic density.  These results are in good agreement with the approximation given in Fig.~\ref{fig:model-xs-mx-gdm} where for the case of $m_X=20$~GeV and $r=1.05$
the correct relic density is given by $g_X=0.252$, independent of the value of $\alpha$.

\begin{figure}[t!]
\begin{center}
\includegraphics[width=0.46\textwidth]{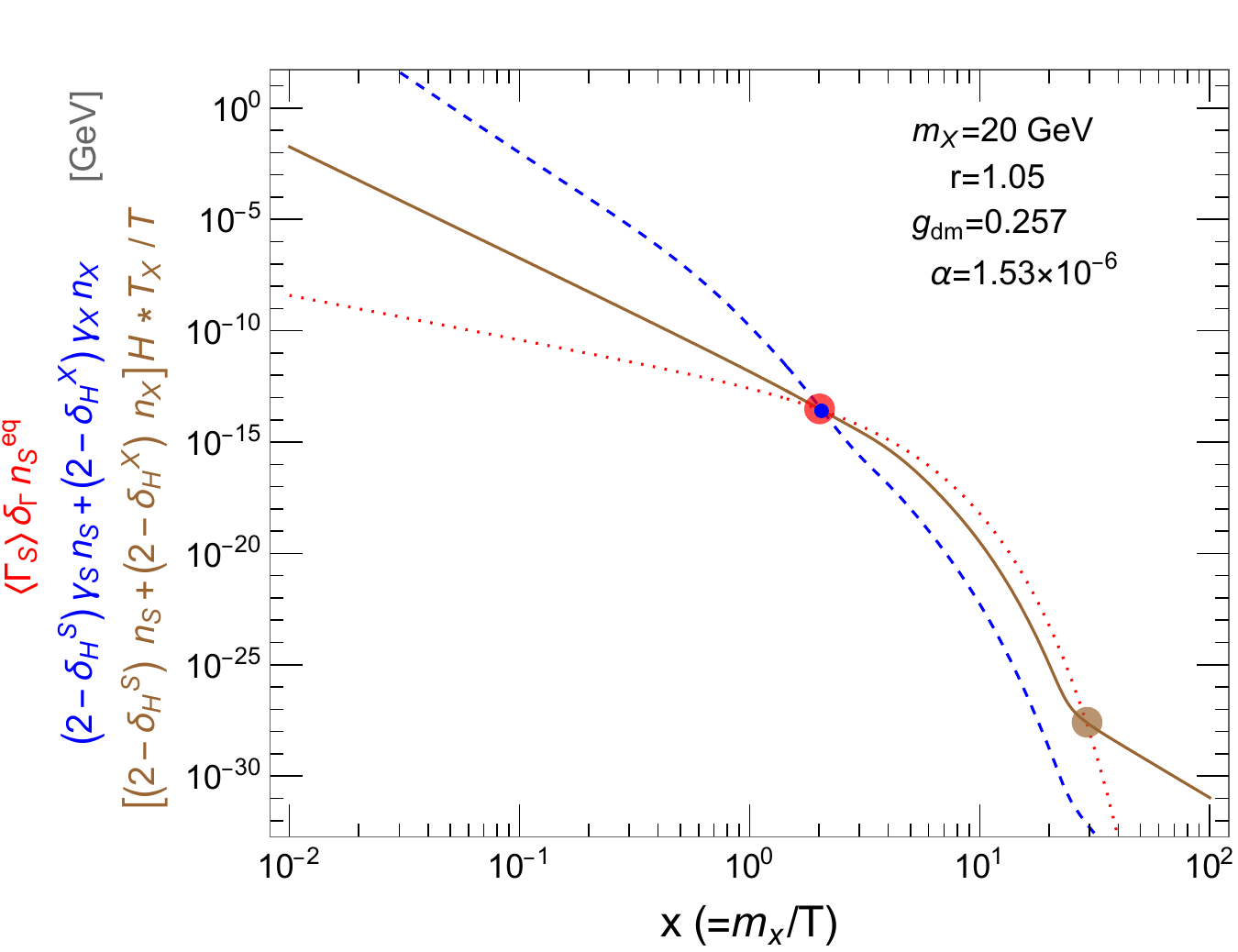}\hskip0.5cm
\includegraphics[width=0.46\textwidth]{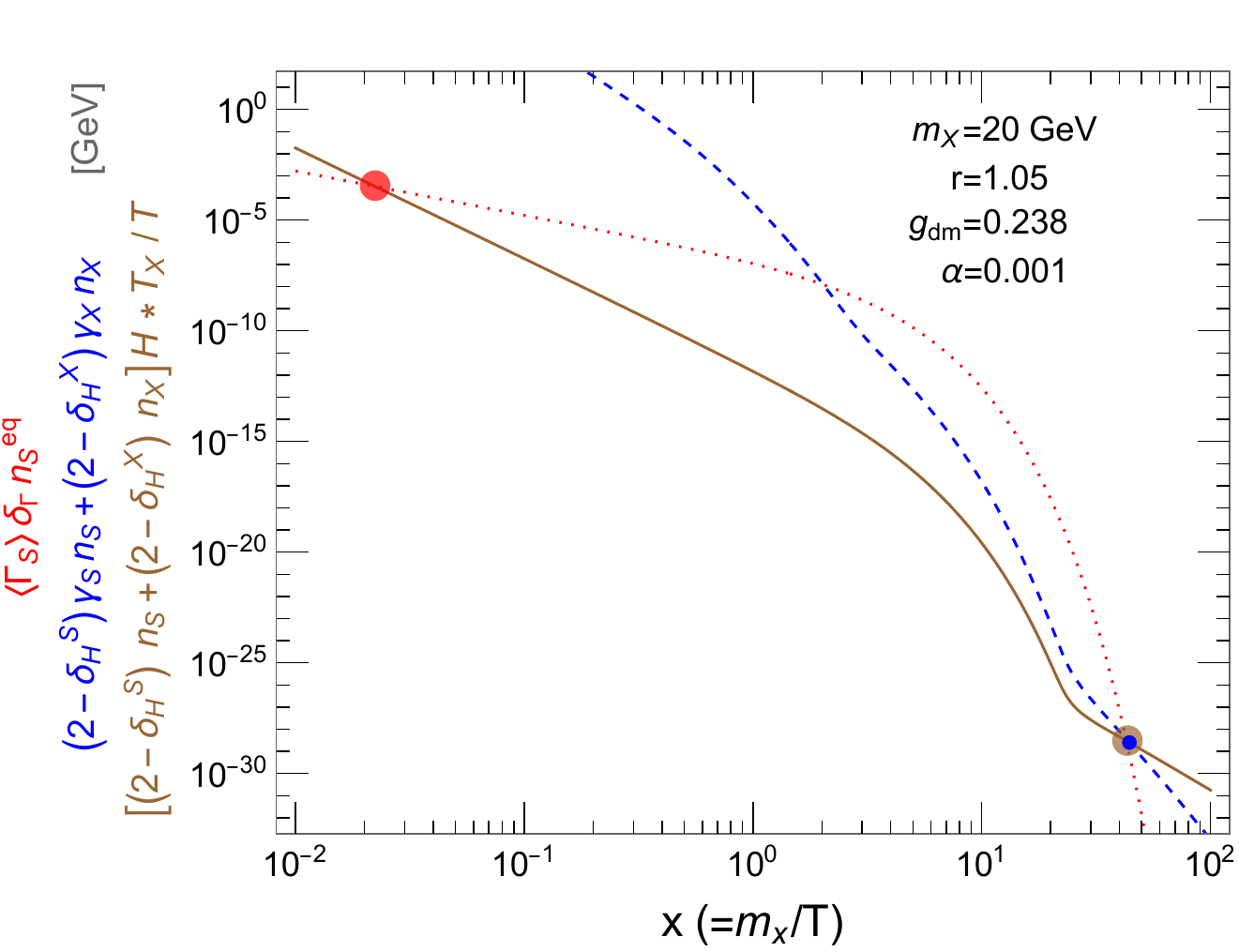} \\
\caption{Comparison of the kinetic energy injection rates from the SM bath to hidden sector via the inverse $S$ decay (red dotted line) and elastic scattering (blue dashed line), and the cooling rate due to the cosmic expansion (brown solid line). Here, these magnitudes of rates have been scaled by a factor of $1/T$.  Below the temperature denoted by the blue dot, the elastic scattering rate falls below the expansion rate in magnitude. In contrast, the inverse $S$ decay rate is larger than the expansion rate when the temperature drops below that, denoted by the red dot. At a later time, the inverse $S$ decay rate falls below the cosmic rate at the temperature denoted by the brown dot.
}
% denotes the temperature below that the hidden sector is kinetically decoupled from the bath and evolves with a different temperature.}
\label{fig:relrate}
\end{center}
\end{figure}

In Fig.~\ref{fig:relrate}, we further illustrate the underlying dynamics of keeping the hidden sector in temperature equilibrium with the bath before kinetic decoupling, where, with a rescaled factor of $1/T$, kinetic energy injection rates from the SM bath to the hidden sector via the inverse $S$ decay (red dotted line) and elastic scattering (blue dashed line), and the cooling rate due to the cosmic expansion (brown solid line) are depicted. 
The red dot (or brown dot) corresponds to $T_{\rm dec}=m_X/x_{\rm dec}$ (or $T_{\rm end}=m_X/x_{\rm end}$), below which we have
$\langle\Gamma_S\rangle \delta_\Gamma n_S^{\rm eq} T >  (\text{or} <) \, [(2 -\delta_H^S) n_S + (2 -\delta_H^X) n_X ] H T_X$, and the blue dot corresponds to $T_{\rm el}=m_X/x_{\rm el}$, above which we have
 $[(2 -\delta_H^S) \gamma_S n_S + (2 -\delta_H^X) \gamma_X n_X ] T >  [(2 -\delta_H^S) n_S + (2 -\delta_H^X) n_X ] H T_X$.
 Here, $\gamma_X$ and $\gamma_S$ are the momentum transfer rates, and their detailed forms are described in Appendix~\ref{app:thermal-temp-elastic}.
The hidden sector receives heat from the bath via elastic scatterings with SM or via the inverse $S$ decay at a rate less than its cooling by cosmic expansion at $x>x_{\rm el}$ or at $x>x_{\rm end}$.

For a lower bound value $\alpha=1.53\times 10^{-6}$ corresponding to $m_X=20$~GeV and $r=1.05$, the hidden sector can be kept in temperature equilibrium with the SM bath mainly due to that the kinetic energy rate transferred from the SM bath to it via elastic scatterings, $S~{\rm SM} \to S~{\rm SM}$ and  $X~{\rm SM} \to X~{\rm SM}$, or via the inverse decay ${\rm SM~SM} \to S$  is larger than its cooling rate by cosmic expansion, depending on whether $x\lesssim 2.05$ or $x\gtrsim 2.05$ (see the left panel of Fig.~\ref{fig:relrate}).

As illustrated in Fig.~\ref{fig:relrate}, for these two cases, $\alpha=1.53\times 10^{-6}$ and $0.001$, the hidden sector is kinetically decoupled from the SM bath well after freeze-out at $x_{k}={\rm max}(x_{\rm el}, x_{\rm end}) \simeq 29$ with $x_{\rm el} \ll x_{\rm end}$ and at $x_{k}={\rm max}(x_{\rm el}, x_{\rm end}) \simeq 43$ with $x_{\rm el} \approx x_{\rm end}$, respectively. See also Table~\ref{tab:list}. These estimates are consistent with numerical results from Boltzmann equations; see the right panel of Fig.~\ref{fig:relic-boltz1-1}.  In Table~\ref{tab:list}, we summarize 
several sets of parameters in the DM mass range from sub-GeV to TeV, resulting in the correct relic density in this forbidden DM model.

\begin{figure}[h!]
\begin{center}
\includegraphics[width=0.55\textwidth]{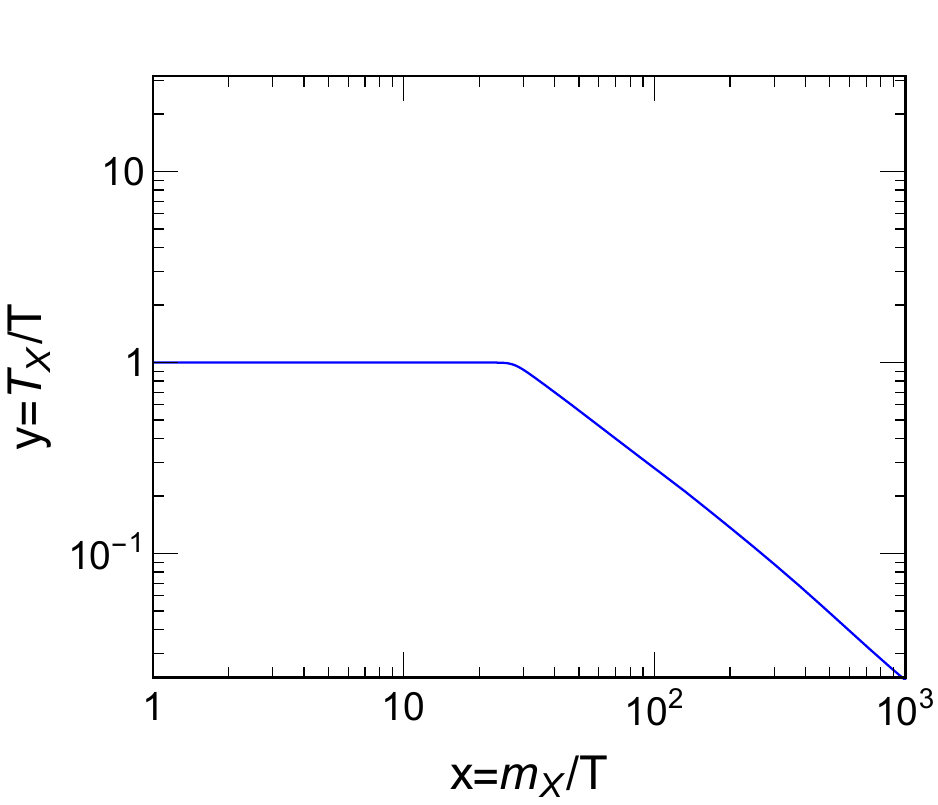}\\
\caption{$T_X/T$ as a function of $x$ for the case that $T_S=T$ can be maintained until after the time of $T_X \not= T$, which is due to a large ${\rm SM~SM} \to S$ rate.
Here we use $m_X=20$~GeV, $r=1.05$, $g_{\rm dm}=0.257$, and $\alpha= 1.53\times 10^{-6}$ as input parameters.}
\label{fig:TX}
\end{center}
\end{figure}

 \begin{table}[h!]
\centering
\begin{tabularx}{0.85\textwidth}{|c|c|>{\centering\arraybackslash}X  |>{\centering\arraybackslash}X| c| c|c|} 
\hline \ \ 
$m_X$(GeV) \ \ & \hspace{0.6cm} $r$  \hspace{0.6cm} & $\alpha$ & \multicolumn{2}{c|}{$g_{\rm dm}$} & \hspace{0.cm}$\text{max}(x_{el}, x_{\rm end})$ \hspace{0.cm} & \hspace{0.15cm} $x_f$ \hspace{0.15cm} \\
\hline
 \hline\multirow{ 7}{*}{0.3} & \multirow{ 2}{*}{1.01}  & $4.68\times 10^{-6}$ & 0.0215   &\multirow{ 2}{*}{(0.0203)}  & 29.8   &\multirow{ 2}{*}{20.4}   \\ \cline{3-4} \cline{6-6}                                                   
                                          &                                   &   0.001                       &  0.0209  &                                          &  41.9  &    \\ \cline{2-7}
                                          & \multirow{ 2}{*}{1.05} & $3.91\times 10^{-6}$ &  0.0349  &\multirow{ 2}{*}{(0.0350)}   & 28.5  &\multirow{ 2}{*}{21.2 }  \\ \cline{3-4}  \cline{6-6}    
                                          &                                   &   0.001                       &  0.0336   &                                         &  40.4  &    \\ \cline{2-7}
                                          & \multirow{ 3}{*}{1.25} &  ${\color{magenta}1.94\times 10^{-6}}$ &    ${\color{magenta}2.52}$
                                          &\multirow{ 3}{*}{(0.387)}    &  ${\color{magenta}18.1}$\footnotemark[4]   & ${\color{magenta} 35.9}$\footnotemark[4] \\ \cline{3-4} \cline{6-7}  
                                          &                                   & $4.82\times 10^{-6}$ &  0.387    &                                          &  24.4  &\multirow{ 2}{*}{22.5}   \\ \cline{3-4} \cline{6-6}  
                                          &                                   &   0.001                       &  0.300    &                                          &  34.1  &   \\ \cline{2-5} 
\hline\multirow{ 7}{*}{20}   & \multirow{ 2}{*}{1.01} & $1.79\times 10^{-6}$ & 0.148      &\multirow{ 2}{*}{(0.139)}    & 30.2   &\multirow{ 2}{*}{22.8}   \\ \cline{3-4} \cline{6-6} 
                                          &                                   &   0.001                       & 0.141      &                                         &  44.5  &    \\ \cline{2-7} 
                                          & \multirow{ 2}{*}{1.05} & $1.53\times 10^{-6}$ &  0.257     &\multirow{ 2}{*}{(0.252)}    & 29.1   &\multirow{ 2}{*}{23.6}   \\ \cline{3-4} \cline{6-6}
                                          &                                   &   0.001                       &  0.238     &                                         &  42.9  &    \\ \cline{2-7}
                                          & \multirow{ 3}{*}{1.25} & $7.88\times 10^{-7}$ &  ---\footnotemark[1]       
                                                                                                                                       &\multirow{ 3}{*}{(3.50)}    & 4.1\footnotemark[3]       &   ---\footnotemark[1]  \\ \cline{3-4}  \cline{6-7}
                                          &                                   &$6.90 \times 10^{-6}$ &  3.50      &                                         & 26.6     &\multirow{ 2}{*}{24.9}   \\ \cline{3-4} \cline{6-6}                                                       
                                          &                                    &   0.001                      &  2.72       &                                        &  86.7\footnotemark[2]   &    \\ \cline{2-7}                                     
\hline\multirow{4}{*}{1000}& \multirow{ 2}{*}{1.01} & $1.20\times10^{-7}$  & 1.10        &\multirow{ 2}{*}{(1.03)}     & 35.2    &\multirow{ 2}{*}{26.6}   \\ \cline{3-4} \cline{6-6}
                                          &                                   &   0.001                      &  1.03       &                                          & 460\footnotemark[2]   &    \\ \cline{2-7}
                                          & \multirow{ 2}{*}{1.05} & $7.65\times 10^{-8}$ &  2.09      &\multirow{ 2}{*}{(2.03)}      & 33.2   &\multirow{ 2}{*}{27.4}   \\ \cline{3-4} \cline{6-6}
                                          &                                    &   0.001                      &  1.92       &                                         & 578\footnotemark[2]   &    \\ \cline{2-7}
                                         \hline
\end{tabularx}
\footnotetext[1]{No solution exists below the perturbative unitarity bound $\sqrt{4\pi}$.}
\footnotetext[2]{$x_{\rm el} = \text{max}(x_{el}, x_{\rm end})$. Except for \footnotemark[2] and \footnotemark[3], $x_{\rm end} = \text{max}(x_{el}, x_{\rm end})$.}
\footnotetext[3]{$x_{\rm dec} \approx x_{\rm end} \approx x_{el}$. This relation is insensitive to the $g_{\rm dm}$ value.}
\footnotetext[4]{For this case, $\text{max}(x_{el}, x_{\rm end})< x_f$. The $x_f$ value is numerically determined when the $SS\to XX$ rate equals the cosmic expansion rate.}
\caption{ \small The coupling $g_{\rm dm}$, which can result in the correct relic density in this forbidden DM model, corresponding to various values of $m_X$ (from sub-GeV to TeV), $r$ and $\alpha$. For comparison, for each set of $m_X$ and $r$, two or three different $\alpha$ are considered; one is a much smaller value corresponding to the lower bound (see the definition in the text), and the largest one in each set is ``0.001". For $r=1.25$, the $\alpha$ value, corresponding to $g_{\rm dm}$ derived from the simplified scenario and resulting in $x_k\equiv \text{max}(x_{el}, x_{\rm end}) >x_f$, is given. Values in the parentheses are the corresponding results of $g_{\rm dm}$ obtained in the simplified scenario of the forbidden DM model and shown in the right panel of Fig.~\ref{fig:model-xs-mx-gdm}.  The freeze-out parameter $x_f$ is the result from the left panel of Fig.~\ref{fig:xf-xs-mx} except for the case with the parameter set, $m_X=0.3$~GeV, $r=1.25$, $\alpha=1.94\times 10^{-6}$, and $g_{\rm dm}=2.52$, for which the hidden sector is kinetically decoupled from the bath before freeze-out.
 }
 \label{tab:list} 
\end{table}

Five remarks are in order as follows.
\begin{enumerate}
\item  The decoupling time of elastic scatterings between $X$ and SM is quite close to that between $S$ and SM.  This can be realized by that the momentum rates $\gamma_X\propto |M_{X {\rm SM}\to X {\rm SM}}|^2/(g_X m_X^3)$ and $\gamma_S \propto  |M_{S {\rm MS} \to S {\rm SM}}|^2/m_S^3$ are of the same order of magnitude. See Appendix~\ref{app:thermal-temp-elastic} for detailed formulas.

\item The kinetic energy rate transferred to the hidden sector via annihilations ${\rm SM~SM} \to SS, XX$ is proportional to $n_{i}^2$, while elastic scatterings $\propto n_{i} n_{\rm SM}$, where $i \equiv X$ or $S$.
While the latter is a $t$-channel dominant process, the former\footnote{For instance, using the detailed balance, the kinetic energy rate transferred to the dark matter is approximately described by $\langle \sigma v \cdot p^2/(3E)\rangle_{{\rm SM~SM} \to XX} (n_{\rm SM}^{\rm eq})^2 = \langle \sigma v \cdot p^2/(3E)\rangle_{XX \to {\rm SM~SM}  } (n_{X}^{\rm eq})^2 \propto (n_{X}^{\rm eq})^2$ for the process with $m_X > m_{\rm SM}$.} is $s$-channel dominant and relatively exponentially suppressed at $T_{\rm el}\lesssim m_X$ due to the nonrelativistic number densities of the hidden sector particles\footnote{Compared with elastic scattering, the annihilations are exponentially suppressed in the temperatures $T\lesssim m_X$. However, they can be enhanced by the $s$-channel resonance in some mass regions. If one chooses the  $\alpha$ lower bound to be $\langle\Gamma_S\rangle \delta_\Gamma n_S^{\rm eq} T = [(2 -\delta_H^S) n_S + (2 -\delta_H^X) n_X ] H T_X$, at $T={\rm max}(T_{\rm el}, T_{\rm ann})$, where $T_{\rm ann}$ is the decoupling temperature of $XX, SS\leftrightarrow {\rm SM~SM} $, it becomes slightly lower in the range $m_X \approx 20\sim 75$~GeV.}. We numerically get $T_{\rm el}\sim  m_X/2$.

\item  
For the scenario that we consider, above the lower bound of $\alpha$, the number densities of the hidden sector particles follow the equilibrium distributions\footnote{For $\alpha =1.53\times 10^{-6}$, although the chemical equilibrium of the hidden sector is still maintained, i.e., $\mu=0$, it shortly evolves with a slightly different temperature, compared with the SM, during $x \approx 3-10$ due to the $3\leftrightarrow 2$ reactions. } before freeze-out.
Here we show that this condition is indeed satisfied. Since $T_{\rm el}\sim  m_X/2$, we estimate the minimum value of $\alpha$, denoted as $\alpha_{\rm fi}$, at which the hidden sector particles will exhibit the equilibrium densities at $x\lesssim 1$.  We consider the hidden sector undergoes a freeze-in process \cite{Hall:2009bx}, where its species have negligible initial densities at $x=x_0 (=m_X/T_0) \to 0$, and the rates of $XX \to {\rm SM~SM}, SS \to {\rm SM~SM}$ and $S\to {\rm SM~SM}$ are much lower than the Hubble expansion rate and thus can be neglected. During freeze-in, we find $Y_S(T)\gtrsim Y_X(T)$. Meanwhile, the hidden sector can quickly reach its internal thermal equilibrium because we consider the case with sizable interactions $XX\leftrightarrow SS$ and $XS\leftrightarrow XS$. We thus use the freeze-in evolution of $Y_S$ to estimate $\alpha_{\rm fi}$.  The evolution equation, rewritten from Eq.~(\ref{eq:boltz-YS}), is given by
\begin{align}
  & \int_0^{Y_S^{\rm eq}(m_X) }dY_S  \nonumber\\
  &\simeq 
   \int_{x_0}^{1} \frac{\tilde{h}_{\rm eff}(T)}{h_{\rm eff}(T)}   \frac{1}{xH} 
    \Bigg[  s \Big( \langle \sigma v \rangle_{SS\to \sum_i {\rm SM}_i {\rm SM}_i} (T) \, ( Y_S^{\text{eq}} (T) )^2 \Big)
    +
   \Gamma_{S}     \frac{K_1(m_S/T)}{K_2( m_S/T)} Y_S^{\text{eq}}(T)
   \Bigg]  dx \,. 
   \label{eq:freezein}
\end{align}
The value of $\alpha_{\rm fi}$ as a function of $m_X$ corresponding to $r=1.01, 1.05, 1.25$ and $1.5$, respectively, has been depicted in Fig.~\ref{fig:alp_mx}.
We find that in the parameter regions above the $\alpha$ lower bound, the number densities of the hidden sector particles can follow the equilibrium distributions.

\item
We have assumed that the $X$ and $S$ particles evolve with the same temperature. This can be correct only when they are simultaneously decoupled from the SM bath. For a lower bound value, {\it e.g.},  $\alpha=1.53\times 10^{-6}$ corresponding to $m_X=20$~GeV and $r=1.05$, although the decoupling of the directly elastic interaction $X~{\rm SM} \leftrightarrow X~{\rm SM}$ occurs earlier at $x\simeq 2.05$ roughly consistent with the blue dot in the left panel of Fig.~\ref{fig:relrate}, the nonrelativistic  $X$ population, however, can be still kept in temperature equilibrium with the bath until after freeze-out  by means of interactions $X~S\leftrightarrow X~S$, where $S$ is in equilibrium with the bath via ${\rm SM~SM} \leftrightarrow S$ until a late time. Here we offer another estimate for the kinetic decoupling temperature of the DM.
We assume that the SM and $S$ remain in kinetic equilibrium with each other, i.e., $T_S=T$,  until after the time of  $T_X\not= T$, resulting from a large ${\rm SM~SM} \to S$ rate. Taking the case of $m_X=20~\text{GeV}, r=1.05, \alpha=1.53\times 10^{-6}$, and $g_X=0.257$ as an example, in Fig.~\ref{fig:TX} we depict $T_X/T$ as a function of $x$, where the decoupling of the kinetic interaction between DM and the bath, described by the interaction $X~S\leftrightarrow X~S$, occurs at $x_k \simeq 29$. 
The temperature evolution formula of the DM, involving the $X~S\leftrightarrow X~S$ elastic scattering, is given in Appendix~\ref{app:x-ed}. 
The result is consistent with what has been shown in Fig.~\ref{fig:relic-boltz1-1}. As for a much larger $\alpha = 0.001$, the resulting $x_k$ is still about 29, which is smaller than that shown in Fig.~\ref{fig:relic-boltz1-1}, but the result for the number density of DM is unchanged.

\item
For a larger $r$, e.g., $r=1.25$, using the lower bound value of $\alpha$ may result in out-of-kinetic equilibrium between the hidden sector and SM bath before freeze-out. For instance, for $m_X$=0.3~GeV, we have $x_k=x_{\rm end}=18.1$ which is less than $x_f=35.9$, while for $m_X$=20~GeV, we get $x_k= x_{\rm end} \approx x_{el} \approx x_{\rm dec}=4.1$ which is much less than the usual $x_f$ value. Here $x_k=x_{\rm end}$ is insensitive to the $g_{\rm dm}$ value.
Therefore, if one wants to have the hidden sector be well in the kinetic equilibrium with the SM bath throughout the DM freeze-out process, a larger $\alpha$, resulting in a larger $S\leftrightarrow \text{SM~SM}$ interaction, is necessary for maintaining the equilibrium. As shown in Table~\ref{tab:list}, using the simplified estimate value, e.g. $g_{\rm dm}=0.387~(3.50)$ for the case of $m_X=0.3$~GeV (20~GeV), we need to increase $\alpha$ from its lower bound $1.94\times 10^{-6}$ ($7.88\times 10^{-7}$) to a higher value  $4.82\times 10^{-6}$ ($6.90\times 10^{-6}$) to have a correct relic density. In summary, for the case of $m_X=0.3$~GeV, we have 
$\alpha \in (4.82\times 10^{-6}, 0.001)$ corresponding to $g_{\rm dm} \in (0.387, 0.300)$, while for the case of $m_X=20$~GeV, we have 
$\alpha \in (6.90\times 10^{-6}, 0.001)$ corresponding to $g_{\rm dm} \in (3.50, 2.72)$.
On the other hand, in addition to increasing the $\alpha$ value, we can instead enlarge the value of $g_{\rm dm}$ so that the DM density is depleted to match its observed relic abundance due to the longer time of interactions between $S$ and $X$. Taking the case with parameters, $m_X=0.3$~GeV, $r=1.25$ and the lower bound value $\alpha=1.94\times 10^{-6}$, as an example, to account for the correct DM relic density, we find that  a much larger $g_{\rm dm}=2.52$ is needed. For this case, when $x \gtrsim x_k (=18.1)$, the hidden sector is kinetically decoupled from the SM bath so that it evolves independent temperature which, featured by the cannibal interactions, logarithmically decreases with the scale factor \cite{Farina:2016llk}. 
As shown in Fig.~\ref{fig:relic-boltz1-out}, the cannibal effect is characterized by the curves in the range $x_k \lesssim x\lesssim x_f$, where $T_X/T>1$ becomes notable. We find that even for the case with $x_k\gg 1$ but $x_k<x_f$, the cannibal interactions are considerably important.
     
 \end{enumerate}

\begin{figure}[t!]
\begin{center}
\vskip-0.35cm
\includegraphics[width=0.457\textwidth]{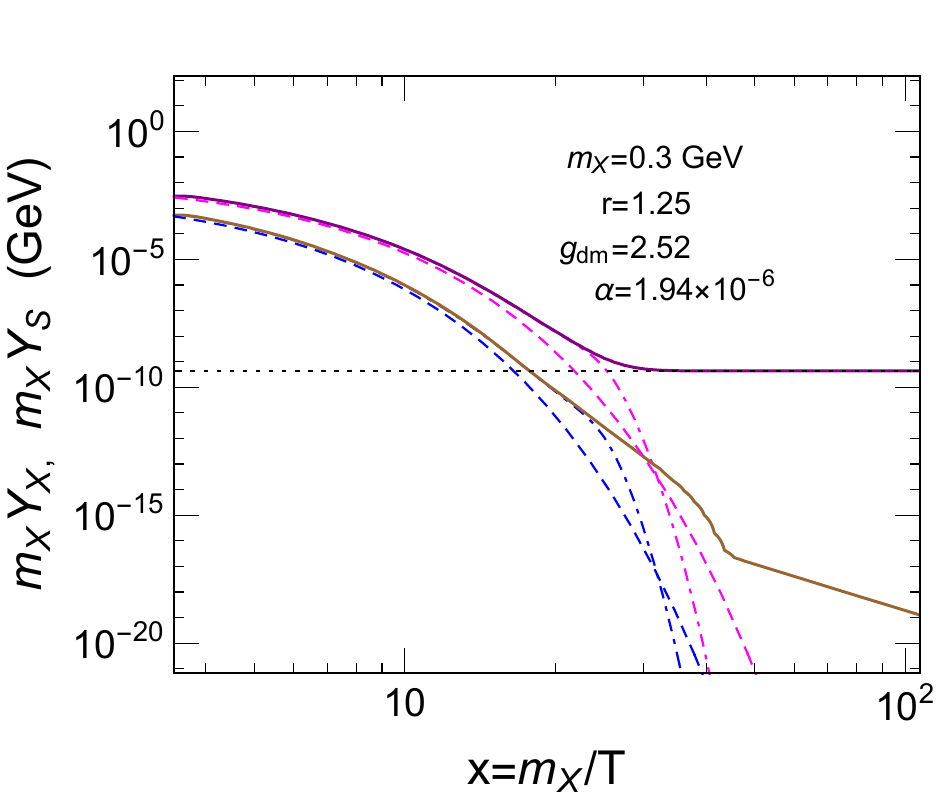}\hskip0.5cm
\includegraphics[width=0.45\textwidth]{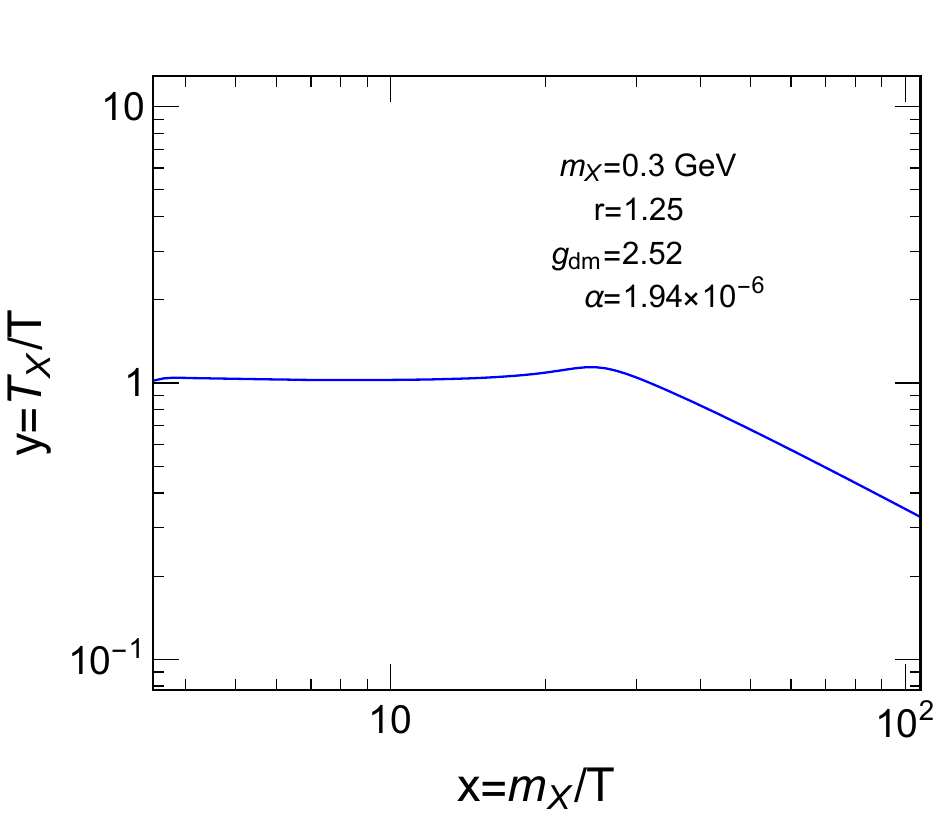}\\
\includegraphics[width=0.457\textwidth]{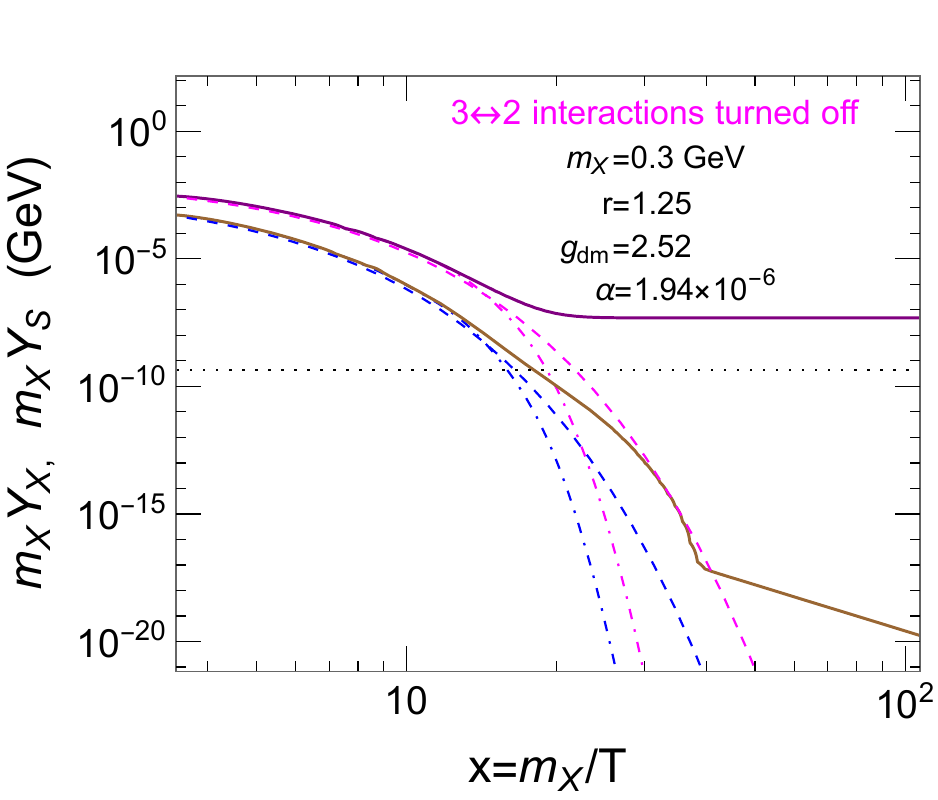}\hskip0.5cm
\includegraphics[width=0.45\textwidth]{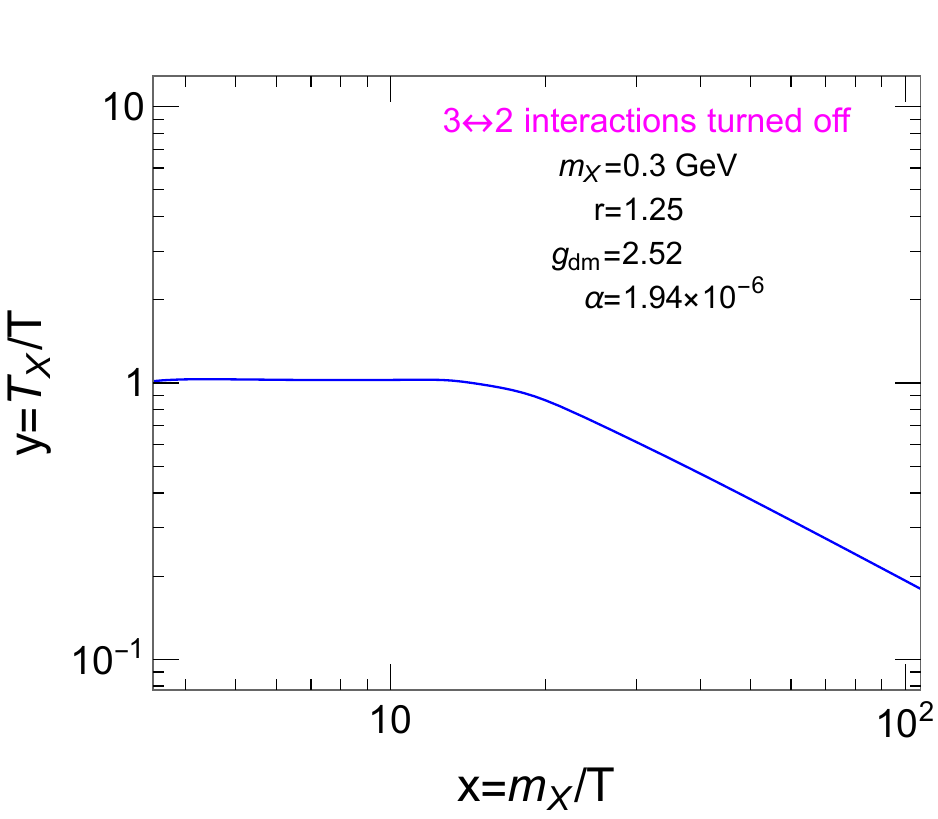}\\
\vskip-0.35cm
\caption{Same as Fig.~\ref{fig:relic-boltz1-1}, but for the case denoted in the plots. The upper ones are in full consideration, while the lower ones are tuned off the cannibal interactions}
\label{fig:relic-boltz1-out}
\end{center}
\end{figure}

\section{Experimental constraints on this forbidden model}\label{sec:constraints}

The DM direct detections can constrain this forbidden DM model.
As this is also a scalar portal model, the scalar mediator can be searched for via decay channels, e.g., $B\to K S$ and $K \to \pi S$ in particle physics experiments. Depending on a slight mixing angle $\alpha$, the $S$ might decay with a displaced vertex or outside the detector. For the latter, it thus plays as an invisible particle. As such, a lower bound of $\alpha$ can be given. 

On the other hand, the big bang nucleosynthesis (BBN) measurement can set a limit on $\alpha$, especially in the case that the $S$ decays out of equilibrium. 
As we will see, the BBN constrains the regions below the black (or blue) solid curve corresponding to $r=1.01$ (or 1.05) in Fig.~\ref{fig:constraints}, where the hidden sector may evolve a temperature different from the SM before $X$ (or $S$) becomes nonrelativistic.  

In Fig.~\ref{fig:constraints}, we display the excluded regions on the $m_X-\alpha$ parameter space set by measurements on the scalar boson $S$, including an estimate of sensitivity projections of  SHiP  \cite{Alekhin:2015byh, SHiP:2018yqc} and NA62 \cite{NA62:2017rwk}, by the direct detection experiment PandaX-4T  \cite{PandaX-4T:2021bab}, including the LZ projected sensitivity \cite{Akerib:2018lyp}, and by the BBN measurements.
Moreover, for comparison, we also show the lower bound of $\alpha$ that can keep the hidden sector in equilibrium with the thermal bath before freeze-out, as discussed in Secs.~\ref{sec:vdm} and \ref{sec:BoltEq}.
For brevity, the limits set by the $S$ measurements are only shown for using $r(=m_S/m_X)=1.01$. However, it can be easy to rescale to the case of $r=1.05$ and others. We do not show the constraint from the invisible Higgs decay since it is weaker than the ones presented here.
The detailed discussions are as follows.

\subsection{Constraints from particle physics experiments}\label{sec:particle}

%\subsubsection{\rm LEP}\label{sec:lep}

At LEP, the L3 collaboration has presented a lower limit of Higgs-like scalar with mass less than 66.7~GeV \cite{L3:1996ome}. 
This experiment has searched for the scalar decaying into the invisible final state through the process $\bar{e} e \to Z^* S$. For $m_S \gtrsim 2 m_\mu$ and $\sin\alpha \gtrsim 0.1$, a light scalar would decay mostly visibly on the detector scale \cite{L3:1996ome,Winkler:2018qyg}. 
For a scalar too heavy to be produced in $B$ meson decays, this offers the most substantial constraint in the $S$ measurements.

%\subsubsection{\rm LHCb}\label{sec:lhcb}

At the current LHCb \cite{LHCb:2015nkv, LHCb:2016awg}, a light scalar $S$ is constrained by the detection of the decay $B\to K^{(*)} \mu^+ \mu^-$ with a scalar resonance in the di-muon channel. For this case, the number of generated events scales with $\sin^4\alpha$. We have used the result given in Fig.~5 of Ref.~\cite{LHCb:2016awg}.

%\subsubsection{\rm BNL-E949}\label{sec:949}

The $K^+ \to \pi^+ \bar{\nu} \nu$ experimental data at BNL-E949 \cite{BNL-E949:2009dza} can be used to interpret the decay model $K^+ \to \pi^+ S$, where the scalar is long-lived and play as an experimentally unobservable particle. The 90\% CL upper limit on ${\rm Br}(K^+ \to \pi^+ S)$  has been set on the mass  of $S$ smaller than $260~{\rm MeV}$, where for the narrow range around the $\pi^0$ mass the limit becomes weaker and has a value $\sim 5.6\times 10^{-8}$, compared with the limits that are less than $ 10^{-10}$ at $m_S =100$ MeV and  $ 4\times 10^{-10}$ at $m_S \sim200$ MeV. 
Moreover, the reinterpretation of the past PS191 experiment, which is related to $K \to \pi S$, can be further used to constrain $m_S$, especially in the range of $110-150$~MeV \cite{Gorbunov:2021ccu}.
Since the number of events of the $K^+ \to \pi^+ S$ mode is proportional to $\sin^2\alpha$, we can thus determine the exclusive region in the $m_S-\alpha$ plane as shown in Fig.~\ref{fig:constraints}. In addition, the NA62 beam dump experiment does a similar search for $K^+ \to \pi^+ + $ missing energy and can thus give an upper limit on $\alpha$ from the measurement of $K^+ \to \pi^+ S$. In Fig.~\ref{fig:constraints}, we depict the region that the future sensitivity of NA62 \cite{NA62:2017rwk} can reach after LHC Run 3, which was estimated in Ref.~\cite{Bondarenko:2019vrb}.

The bean dump experiments can efficiently detect long-lived light scalars, where detectors are situated far from the fixed target. In Fig.~\ref{fig:constraints}, we display the limit of  CHARM \cite{CHARM:1985anb} and the projected sensitivity of SHiP \cite{Alekhin:2015byh, SHiP:2018yqc}, both of which were analyzed by Winkler  \cite{Winkler:2018qyg}.

Across the parameter space shown in Fig.~\ref{fig:constraints}, for $m_S\lesssim 5$~GeV the region above the black (or blue) solid curve corresponding to $r=1.01$ (or 1.05), where the hidden sector is in thermal equilibrium with the SM before freeze-out, is considerably excluded by the current measurements
\footnote{Especially,  the present constraints can exclude the entire $m_S\lesssim 260$~MeV parameter space, where $\alpha$ is large enough to keep the hidden sector in thermal equilibrium before freeze-out.} and further constrained by future experiments.

\subsection{Constraints from direct detections}\label{sec:direct}

\begin{figure}[t!]
\begin{center}
\includegraphics[width=0.55\textwidth]{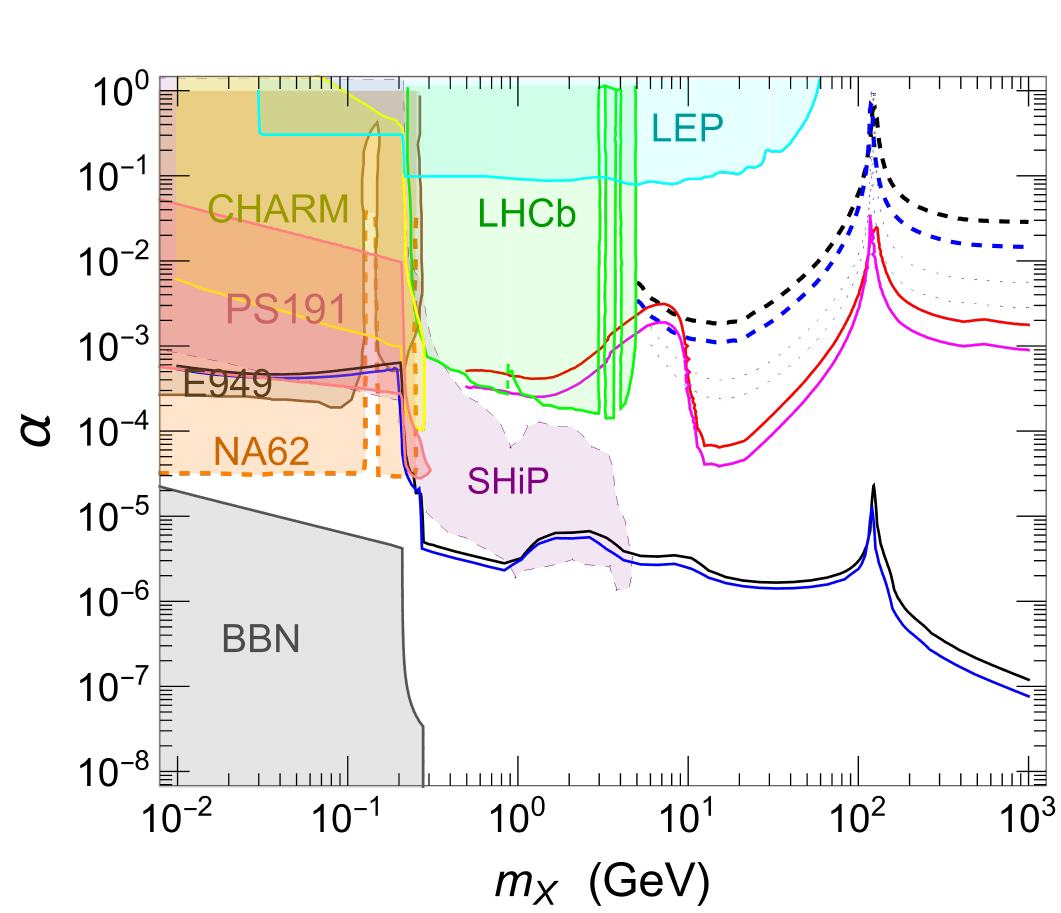}
\caption{Constraints on the $m_X-\alpha$ parameter space. (1) From the hidden scalar search, colored regions with solid boundaries are excluded by experiments denoted in the plot, while colored regions with dashed boundaries indicate sensitivity projections. For brevity, $r (=m_S/m_X)=1.01$ is used.
(2)  With $r=1.01$ (black) and 1.05 (blue), the PandaX-4T~\cite{PandaX-4T:2021bab} bound and LZ projected sensitivity \cite{Akerib:2018lyp} are respectively shown by the dashed and dotted lines, 
corresponding to use of the local DM density $\rho_\odot =$ 0.4 $\text{GeV/cm}^3$.
The SM neutrino floor \cite{Billard:2013qya} is denoted by the red (or magenta)  solid curve corresponding to $r=1.01$ (or 1.05).
For comparison, the lower bounds of $\alpha$, keeping the hidden sector in equilibrium with the thermal bath before freeze-out, with $r=1.01$ (black solid curve) and $1.05$ (blue solid curve) are shown.}
\label{fig:constraints}
\end{center}
\end{figure}

The direct detection experiments can set constraints on couplings of the present forbidden DM model, especially in the range $m_X\gtrsim 5$~GeV.
The dominant constraint is from the measurement of the spin-independent cross section, which results from
the elastic $t$-channel DM-nucleon interactions mediated by $S$ and $h$, given by
\begin{align}
\sigma_N =  \frac{\mu_{XN}^2 m_N^2 f_N^2 g_{\rm dm}^2}{4\pi} \frac{\sin^2 2\alpha}{v_H^2} \left( \frac{1}{m_S^2} - \frac{1}{m_h^2} \right)^2 \,,
\end{align}
where $\mu_{XN}$ is the reduced mass of $X$ and nucleon $N$, and the effective coupling $f_N \simeq 0.3$  (see Ref.~\cite{Yang:2020vxl} and references therein).

Since for a small $r\lesssim 1.05$ the coupling $g_{\rm dm}$ is very weakly dependent on the variation of $\alpha$ in accounting for the correct relic density in this forbidden model,
 for simplicity, we use the value of $g_{\rm dm}$ shown in Fig.~\ref{fig:model-xs-mx-gdm} as an input, which is a good approximation compared with the true one, where differences reside within a factor of 7\%  (see Table~I for comparison).
 Thus, the direct detections can directly constrain the value of $\alpha$.
In Fig.~\ref{fig:constraints}, we show the 90\% C.L. bounds on the mixing angle $\alpha$ from PandaX-4T \cite{PandaX-4T:2021bab} and LZ projected sensitivity \cite{Akerib:2018lyp}, together with the SM neutrino floor  \cite{Billard:2013qya}. The current PandaX-4T bound on the DM has reached the neutrino floor at $m_X \approx 7\sim 9$~GeV. In the region $m_X>5$~GeV, more restrictive constraints on the parameter space of the forbidden DM model from direct searches can be considerably improved in the future.

\subsection{Big bang nucleosynthesis}\label{sec:bbn}

 The BBN significantly constrains the process that the hidden scalar $S$ has a longer lifetime and decays out-of-equilibrium during and after DM freeze-out.
As such, we can limit the mixing angle $\alpha$.  The Universe might experience reheating due to the entropy production by the out-of-equilibrium decay of $S$. Moreover, if the reheating temperature is close to the neutrino decoupling temperature, the relative ratios of light element abundances will be changed.
In Fig.~\ref{fig:constraints}, we identify the excluded region with the lifetime $\tau_S >1$~sec and corresponding to $r=1.01$. This lifetime limit is usually adopted and consistent with the value given in Ref.~\cite{Kawasaki:2000en}. Note that, 
 for the decay channels in the non-perturbative QCD region corresponding to $2m_\pi \leq m_S \leq 2$~GeV, we model the hadronic final-state modes
 by extrapolating $gg, ggg$, and $g \bar{q} q$ modes in the NLO perturbative spectator model, where the renormalization scale is chosen to be $\mu= m_S/1.76$ for 1~GeV~$\leq \sqrt{s} \leq 2$~GeV and 
 $\mu =0.57$~GeV for  $2 m_\pi \leq m_S \leq 1$~GeV.  The resulting width is approximately consistent with Fig.~4 in Ref.~\cite{Winkler:2018qyg}.
  See also the discussions after Eq.~(\ref{app:sgg}) for the decay width of $S$.
  
As shown in Fig.~\ref{fig:constraints}, the BBN mainly constrains the regions, which correspond to the process that the hidden sector is kinetically decoupled from the SM bath at temperature $T\gtrsim m_{S,X}$. For regions of the parameter space below the black (or blue) solid curve, the evolution of the hidden sector may undergo a freeze-in process, which is beyond the scope of this paper.

\section{Summary}\label{sec:conclusions}

The forbidden dark matter is described by its annihilations into heavier states, which vanishes at zero temperature limit, but occurs at finite temperatures in the early Universe.
For the forbidden annihilation channel, if the final state is scalar mediators that couple to SM matter through mixing with the SM Higgs, the resulting signals from DM and the mediator could be detected by direct detections and particle physics experiments.

We have explored the secluded vector DM model in the mass range from sub-GeV to TeV to provide a concrete realization of this forbidden DM scenario.
This DM model is UV-complete. It is the minimum extension of the Standard Model, where the vector DM is coupled to the extended singlet $S$ in the hidden sector.   Through mixing with the SM Higgs, searches for the mediator decay signals, including displaced vertices or missing energy, could be possible at colliders or fixed target experiments. 

We have considered that the hidden sector is in thermal equilibrium with the SM bath before freeze-out.
In this forbidden dark matter scenario,  the depletion of dark matter following the Boltzmann suppression is dominated by annihilations into two heavier but unstable scalars, $XX \to SS$, with mass $m_X < m_S$. In Sec.~\ref{thermal-freezeout}, we have given generic estimates for the values of freeze-out temperature parameter $x_f$ and the thermally averaged $SS\to XX$ annihilation cross section. Considering a simplified scenario, we have further made a model estimate for the $X$-$S$ coupling and the mixing angle of the SM Higgs and hidden scalar in Sec.~\ref{sec:vdm}.  The former ($g_{\rm dm}$), depending on $m_X$ and $r (=m_S/m_X)$, can be determined by having the correct relic density and constrained by the perturbative unitarity bound, while for the latter ($\alpha$), the lower bound of the region that can account for the temperature equilibrium between the hidden sector and the SM before freeze-out
is set by the condition that the heating rate transferred to the hidden sector via elastic scatterings or inverse $S$ decay is larger than the cooling rate by cosmic expansion.

We solve the coupled Boltzmann equations for the freeze-out forbidden dark matter model scenario, including the yields ($Y_X$, $Y_S$) and temperature ratio ($T_X/T$). To illustrate the underlying dynamics of the temperature evolution, we have shown the individual evolutions for the heating rates from elastic scatterings or inverse $S$ decays and the cooling rate due to cosmic expansion. The dynamics of this forbidden DM model depends on four parameters, $m_X, r(=m_S/m_X), g_{\rm dm}$ and $\alpha$. We have found that the value of $g_{\rm dm}$ accounting for the correct relic density is mainly a function of $m_X$ and $r$ and is close to the result derived in the simplified scenario (see Sec.~\ref{thermal-freezeout}) in the case of a small $r$. Moreover, it is weakly dependent on $\alpha$ in most regions of the parameter space where the hidden sector and SM are in thermal equilibrium with each other until after freeze-out.   We have shown that a mass ratio as large as, e.g., $r=1.5$, is allowed only in the sub-GeV region (see Fig.~\ref{fig:model-xs-mx-gdm}). Furthermore, imposing the perturbative unitarity bound $\sqrt{4\pi}$ on the coupling $g_{\rm dm}$, we have found that the mass splitting for $m_{X,S} \gtrsim O({\rm TeV})$ is required to be within the percent level  (see also Fig.~\ref{fig:model-xs-mx-gdm}).

However, for the case with a larger $r$, e.g., $r=1.25$ and a mixing angle $\alpha$ close to its lower bound value, the hidden sector may be kinetically decoupled from the SM bath before freeze-out. In order for the hidden sector to be well in the kinetic equilibrium with the SM bath throughout the DM freeze-out process, we need to have a larger $\alpha$ to guarantee the equilibrium through a more significant $S\leftrightarrow \text{SM~SM}$ interaction,
On the other hand, instead of increasing the $\alpha$ value, we can enlarge $g_{\rm dm}$ to have a longer time of interactions between $S$ and $X$ so that the DM density is depleted to match its observed relic abundance. If doing so, the cannibal effect is significant in the range $x_k \lesssim x\lesssim x_f$, where $T_X/T>1$ becomes remarkable (see Fig.~\ref{fig:relic-boltz1-out}).

This forbidden DM model can be constrained by DM direct detections and by mediator searches at colliders and fixed target experiments.
In the lower mass range of $m_S \lesssim 5$~GeV, the scalar mediator can be produced in rare $B, D$ meson decays. Depending on the mixing angle $\alpha$, a long-lived $S$ might decay with a displaced vertex or escape the detector as missing energy.  In this mass range, the current experimental constraints and projected sensitivities can cover the most parameter space above the lower bound of $\alpha$ defined in Sec.~\ref{thermal-freezeout}.
In the region $m_X>5$~GeV, 
more restrictive constraints on the parameter space from direct searches can be considerably improved in the future.
However, a large part of the parameter space still survives the projected constraint from LZ.

\acknowledgments \vspace*{-1ex}
 This work was partly supported by the National Center for Theoretical Sciences and the National Science and Technology Council of Taiwan under grant numbers MOST 110-2112-M-033-004 and MOST 111-2112-M-033-006.

\appendix

\section{The $2 \to 2$ annihilation cross sections}\label{app:2to2}
 
 Through this paper, we use the notation $\langle \sigma v \rangle \equiv \langle \sigma v_{\text{M\o l}}\rangle$, where $v_{\text{M\o l}}$ is the M{\o}ller velocity. 
 For $i\, i\to j\, j$, the thermally averaged annihilation cross section $\langle \sigma v \rangle_{ii\to jj} (T_k)$ at temperature $T_k$ is described by \cite{Gondolo:1990dk}
\begin{align}
\langle \sigma v  \rangle_{ii\to jj} (T_k) =& \frac{1}{8m_i^4 T_k K_2^2 (m_i/T_k)}  \nonumber\\
 & \times \int_{4m_i^2}^{\infty}  \theta(s-4 m_j^2)
(\sigma v_{\text{lab}} )_{ii\to jj}  ( s-2 m_i^2)  (s-4m_i^2)^{1/2} K_1 (\sqrt{s}/T_k) ds, \label{eq:thermal-average}
\end{align}
where $K_{1,2}$ is the modified Bessel functions, $m_i$ is the mass of the initial particle ``$i$",   $s$ is the invariant mass squared of the two incoming particles, and  $ v_{\rm lab} = [s (s-4m_X)]^{1/2} / (s-2m_X) $ is the lab velocity calculated in the rest frame of one of the incoming particles.  

\subsection{The  $SS\to XX$ annihilation}\label{app:ss2xx}

For the $SS\to XX$ annihilation via the 4-vertex, s-channel, and $u$- and $t$- channels, we have
\begin{align}
& \!\!\!  (\sigma v_\text{lab})_{SS\to XX}
= \frac{g_{\rm dm}^2  c_\alpha^2 \sqrt{s-4 m_X^2}}{32 \pi  m_X^4 \sqrt{s}
   \left(s-2 m_S^2\right) \left(\Gamma_S^2 m_S^2+\left(m_S^2-s\right)^2\right)}
\nonumber\\
&\times  \Bigg\{
  - 
      \frac{8 g_{\rm dm} {c_\alpha}\coth ^{-1}
      \left(\frac{2 m_S^2-s}{\sqrt{s-4 m_S^2} \sqrt{s-4 m_X^2}}
      \right) }{\left(2 m_S^2-s\right) \sqrt{s-4 m_S^2} \sqrt{s-4 m_X^2}}
          \Bigg[
             g_{\rm dm} {c_\alpha} \left(\Gamma_S^2m_S^2+ \left(m_S^2-s\right)^2 \right)
             \nonumber\\
 &\ \  \ \ \times  \left(3 m_S^8-2 m_S^6  (6 m_X^2+s)
    +2 m_S^4 \left(4m_X^4+5 m_X^2 s\right) - 4 m_S^2 m_X^2 \left(4 m_X^4+s^2\right)
    +24 m_X^6 \left(s-2m_X^2\right)
   \right)
   \nonumber\\
 &\ \  \ \  -g_{SSS} m_X  (m_S^2-s)  (2 m_S^2-s) 
  \left(m_S^2 (m_S^2  -4  m_X^2) (2 m_X^2+s)
  +2 m_X^2 (12 m_X^4-2 m_X^2 s+s^2) \right)
   \Bigg]
   \nonumber\\
&  +\frac{2 g_{\rm dm}^2 c_\alpha^2  \left(\Gamma_S^2 m_S^2+ (m_S^2-s)^2\right) }{m_S^4-4 m_S^2 m_X^2+m_X^2 s}
 \nonumber\\
& \ \  \times \left(3 m_S^8-20 m_S^6
   m_X^2+m_S^4 (46 m_X^4+6 m_X^2 s)  -4 m_S^2  (14 m_X^6+5 m_X^4 s) +48
   m_X^8+6 m_X^6 s+4 m_X^4 s^2\right)
   \nonumber\\
 &  + g_{SSS}^2 m_X^2
   (12 m_X^4-4 m_X^2 s+s^2) -4 g_{SSS} g_{\rm dm} m_X {c_\alpha}  (m_S^2-s)
    \left(m_S^2 (2 m_X^2+s) -m_X^2  (6 m_X^2+s) \right)
\Bigg\}
\,,
\end{align}  
 where $s_\alpha \equiv \sin\alpha, c_\alpha \equiv \cos\alpha$ and
 \begin{align}
g_{SSS}=  -\frac{ 3 c_\alpha^3  m_S^2}{v_S} + \frac{3 s_\alpha^3 m_S^2}{v_{H}} \,. \label{eq:gsss}
\end{align}
Here we have neglected the $s$-channel annihilation via a virtual SM Higgs, $SS \to h^* \to XX$, which is suppressed by $\sin^2\alpha$ in the amplitude level.

\subsection{$XX \to \text{SM~SM}$ and $SS \to \text{SM~SM}$}\label{app:hidden2SM} 

For the number density evolutions of the hidden sector particles $X$ and $S$, their annihilation cross sections into the SM are relevant to the chemical equilibrium between them and the SM.  

In the higher energy region of the annihilation with the invariant mass larger than 2~GeV and the lower energy region with $\sqrt{s}< 2m_\pi$, where for the latter, the final states are dominated by pure leptonic modes, we employ the perturbative spectator model to calculate the annihilations, i.e., the final state is two elementary particles in SM.
The kinematically allowed processes, where the $u$- and $t$-channels, if existing, are further suppressed by $\sin\alpha$ in the amplitude level, are dominated by the $s$-channel annihilation mediated by a virtual $S$ or $h$.
The relevant annihilations are $XX, SS \to \bar{q} q, \bar{\ell} \ell, gg, WW, ZZ, hh$ with $q\equiv$ quark and $\ell\equiv$ lepton, where the cross sections can be approximately written as
\begin{align}
  ( \sigma v_\text{lab})_{XX\to ff}
  = & \frac{s^{1/2} }{9(s- 2 m_X^2)}
  \sum_{\lambda_1, \lambda_2} (\epsilon_{X}^{\lambda_1}\cdot  \epsilon_{X}^{\lambda_2})(\epsilon_{X}^{\lambda_1*}\cdot  \epsilon_{X}^{\lambda_2*}) \nonumber\\
     &\times  \bigg[ \Gamma_S \bigg]_{m_S \to \sqrt{s}} ^{S\to ff}
  \bigg| \frac{g_{XXS} \tilde{g}_{Sff}}{s-m_S^2 +i m_S\Gamma_S} + \frac{g_{XXh} \tilde{g}_{hff}}{s-m_h^2 +i m_h\Gamma_h} \bigg|^2 \,,
\end{align}
and
\begin{align}
 ( \sigma v_\text{lab})_{SS\to ff}
  =& \frac{s^{1/2} }{s- 2m_S^2} \bigg[ \Gamma_S \bigg]_{m_S \to \sqrt{s}} ^{S\to ff}
  \bigg| \frac{g_{SSS} \tilde{g}_{Sff}}{s-m_S^2 +i m_S\Gamma_S} + \frac{g_{SSh} \tilde{g}_{hff}}{s-m_h^2 +i m_h\Gamma_h} \bigg|^2  \,,
\end{align}
where $[\Gamma_S]^{S\to ff}_{m_S\to \sqrt{s}}$  means the $S\to ff$ partial width with the invariant mass replacing $m_S$ and with a mixing angle $\alpha=\delta$ fixed so that $g_{Sff}(\alpha=\delta) \not= 0$, ``$f f$" corresponds to the kinematically opened SM pair, $\epsilon_{X}^{\lambda_1}$ and $\epsilon_{X}^{\lambda_2}$ are the polarizations of the two incoming $X$'s, and 
\begin{align}
 & g_{XXS} \equiv   2 g_{\rm dm} m_X c_\alpha\,, \hskip2.4cm
    g_{XXh} \equiv 2 g_{\rm dm} m_X s_\alpha\,, \\
 & \tilde{g}_{Sff} =  \frac{g_{Sff}(\alpha)}{g_{Sff}({\alpha =\delta})\big|_{m_S \to \sqrt{s}} } \,,  \hskip0.98cm
    \tilde{g}_{hff} =    \frac{g_{hff}(\alpha)}{g_{Sff}({\alpha =\delta})\big|_{m_S \to \sqrt{s}} } \,, \\
&   g_{SSh}=
-  \frac{c_\alpha^2 s_\alpha (2 m_S^2 + m_h^2)}{v_S} - \frac{c_\alpha s_\alpha^2 (2 m_S^2 + m_h^2)}{v_{H}} \,,  \label{eq:ghss}
\end{align}
with
\begin{align}
g_{S\bar{q}q}            =& -\frac{m_q}{v_H} s_\alpha \,,   \hskip1cm
g_{h\bar{q}q} =  \frac{m_q}{v_H} c_\alpha \,, 
\label{app:coup-sqq}\\
g_{S\bar{\ell}\ell}            =& - \frac{m_\ell}{v_H} s_\alpha \,,   \hskip1cm
g_{h\bar{\ell}\ell} =  \frac{m_\ell}{v_H} c_\alpha \,, \\
g_{S VV} = & -\frac{2 m_V}{v_H} s_\alpha \,,   \hskip0.5cm
g_{h VV} = \frac{2 m_V}{v_H} c_\alpha \,, \hskip1cm  \text{with~} V\equiv W, Z \,, \\
g_{Shh}   = &   \frac{c_\alpha^2 s_\alpha (2 m_h^2 + m_S^2)}{v_H} - \frac{c_\alpha s_\alpha^2 (2 m_h^2 + m_S^2)}{v_{S}}\,,   \hskip0.2cm
g_{hhh}   = -\frac{ 3 c_\alpha^3  m_h^2}{v_H} + \frac{3 s_\alpha^3 m_h^2}{v_{S}} \,, \label{app:ghhh} \\
g_{Sgg}  = & - \frac{\alpha_s}{4 \pi}\frac{s_\alpha}{v_H} \sum_{q=t,b} \tau_q^S [1-(1-\tau_q^S) f(\tau_q^S)]  \,,   \hskip0cm
g_{hgg}  =   \frac{\alpha_s}{4 \pi}\frac{c_\alpha }{v_H} \sum_{q=t,b} \tau_q^h [1-(1-\tau_q^h) f(\tau_q^h)] \,. 
\label{app:coup-sgg}
\end{align}
Here
\begin{align}
f(\tau_q^i )  
&=  
\left\{ 
   \begin{array}{lr} 
        \Big(\arcsin \sqrt{1/\tau_q^i} \Big)^2
        \, , \ \ & \text{for}\quad  \tau_q^i  \geq 1 
      \\ 
      -\frac{1}{4} \left[ \log \frac{1+\sqrt{1-\tau_q^i }}{1-\sqrt{1-\tau_q^i }} - i \pi\right]^2 
        \, , & \text{for}\quad \tau_q^i  < 1 
   \end{array} 
\label{app:gtau} 
\right. \;,
\end{align}
with $\tau_q^S =\tau_q^h = 4 m_q^2 / s$. The effective $hgg$ and $Sgg$ couplings are given by the Lagrangian (see Ref.~\cite{Buschmann:2014sia} and references therein),
\begin{equation}
{\cal L}\supset \frac{1}{2} (g_{hgg}  h + g_{Sgg} S) G^{\mu\nu} G_{\mu\nu} \,. \label{app:sgg}
\end{equation}
$\Gamma_S$ and $\Gamma_h$ are the decay widths of the hidden Higgs and SM Higgs, and their detailed results can be found in Refs.~\cite{Yang:2019bvg, Yang:2020vxl, Djouadi:2005gi, Djouadi:2005gj}.    
Two remarks are in order. First, in the calculation, we have included the final states, $WW^*, ZZ^*$, and $h h^*$, where the particle with a superscript star is off-shell. These modes correspond to three-body final states or multi-body final states in annihilation. Second, for the $gg$ mode, we further include the QCD NLO corrections. To this order, the $ggg$ and $q \bar{q} q$ need to be taken into account consistently.
Thus, involving  $gg, ggg$ and $q \bar{q} q$ final-state modes, we effectively take the replacement $g_{igg}^{\rm NLO} =\big( 1+E(m_i)\alpha_s/\pi \big)^{1/2} g_{igg}^{\rm LO} $ \cite{Spira:1995rr} with $i\equiv S, h$ and
 \begin{equation}
 E(m_i) =  \frac{95}{4} - \frac{7}{6} N_f +\frac{33-2N_f}{6} \log \frac{\mu}{m_i} \,,
 \end{equation}
 where $N_f$ is the number of active flavors, and $\mu$ is the renormalization scale. In the higher energy region $\sqrt{s} > 2$~GeV, we use $\mu= \sqrt{s}/1.76$.
 
 In the non-perturbative QCD region $2m_\pi \leq \sqrt{s} \leq 2$~GeV,  we extrapolate $gg, ggg, g \bar{q} q$ modes in the NLO perturbative spectator model to model the hadronic final-state modes by choosing $\mu= \sqrt{s}/1.76$ for 1~GeV~$\leq \sqrt{s} \leq 2$~GeV and 
 $\mu =0.57$~GeV for  $2\pi \leq \sqrt{s} \leq 1$~GeV. The result is approximately consistent with Fig.~4 in Ref.~\cite{Winkler:2018qyg}.

\section{The terms of $3 \leftrightarrow 2$ number-changing interactions in Boltzmann equations of the number densities}\label{app:boltz1}

During the time that the $3 \leftrightarrow 2$ cannibal process of the hidden sector is active, the $X$ and $S$ populations can keep well in temperature equilibrium with each other through $X S \leftrightarrow X S$. Therefore, 
for the terms of $3 \leftrightarrow 2$ number-changing interactions in Eqs.~(\ref{eq:boltz-1}) and (\ref{eq:boltz-2}),
we can set $T_X=T_S$ and have
\begin{align}
 \{3 \leftrightarrow 2 \}_X   =  &
  \frac{\langle \sigma v^2 \rangle_{SSS\to XX}}{3} 
    \bigg( n_S^3 -n_X^2 \frac{(n_S^{\text{eq}} (T_X))^3}{(n_X^{\text{eq}} (T_X))^2} 
    \bigg)
     -  \frac{\langle \sigma v^2 \rangle_{XXX\to XS}}{3} \bigg( n_X^3 -n_X n_S \frac{(n_X^{\text{eq}}(T_X))^2}{n_S^{\text{eq}}(T_X)} \bigg)   
    \nonumber\\
    &  -  \langle \sigma v^2 \rangle_{XXS\to SS} 
    \bigg( n_X^2 n_S -n_S^2  \frac{(n_X^{\text{eq}} (T_X))^2}{(n_S^{\text{eq}} (T_X))} 
    \bigg)  \,,
     \label{app:3-2-x} 
%\\
\end{align}
\begin{align}
 \{3 \leftrightarrow 2 \}_S    =   
 &   \frac{ \langle \sigma v^2 \rangle_{XXX\to XS}}{6} 
     \bigg( n_X^3 -n_X n_S \frac{(n_X^{\text{eq}}(T_X))^2}{n_S^{\text{eq}}(T_X)} \bigg) 
     - \frac{ \langle \sigma v^2 \rangle_{XSS\to XS}}{2} 
     \Big( n_X n_S^2 - n_X n_S n_S^{\text{eq}}(T_X)   \Big) 
      \nonumber\\
 &  - \frac{ \langle \sigma v^2 \rangle_{SSS\to XX}}{2}
  \bigg( n_S^3 -  n_X^2  \frac{(n_S^{\text{eq}} (T_X) )^3}{(n_X^{\text{eq}} (T_X))^2} \bigg)
  - \frac{ \langle \sigma v^2 \rangle_{SSS\to SS} }{6} \Big( n_S^3 -n_S^2 n_S^{\text{eq}}(T_X)   \Big) \nonumber\\ 
 & + \frac{ \langle \sigma v^2 \rangle_{XXS\to SS} }{2}
     \bigg( n_X^2 n_S - n_S^2  \frac{(n_X^{\text{eq}}(T_X))^2 }{n_S^{\text{eq}}(T_X)} \bigg)
   \,. \label{app:3-2-s}
\end{align}
For the calculation of $3\to 2$ processes, we approximate the three initial particles in the zero velocity limit, i.e.,  $\langle \sigma v^2 \rangle_{3\to2} \approx \sigma v^2 {}_{3\to2}$.
The detailed expressions of $\langle \sigma v^2 \rangle_{3\to2} $ can be found in Appendix~D of Ref.~\cite{Yang:2019bvg}.

\section{ The collision terms in the temperature Boltzmann equations }\label{app:col-T}

In general, in the second moment Boltzmann equation related to the temperature,
 the collision terms, due to the interaction  ``$a_1 \cdots b_1 \cdots \leftrightarrow a_1' \cdots b_1' \cdots $'',
can be written in the form,
\begin{align}
 g_{a_1} & \int   \frac{d^3p_{a_1}}{(2\pi)^3}   C \Big[f_{a_1}\cdot \frac{ {\bf p}_{a_1}^2}{3 E_{a_1} }  \Big] 
  =   \int  d\Pi_{a_1}  \cdots  d\Pi_{b_1} \cdots
       d\Pi_{a_1'}  \cdots    d\Pi_{b_1'}  \cdots 
     \nonumber\\
& \times (2\pi)^4 \delta^{(4)} ( p_{a_1} + \cdots  p_{b_1} + \cdots - p_{a_1'}  - \cdots  - p_{b_1'} - \cdots )
|M|^2  
\frac{1}{S\, S_{\rm in}\, S_{\rm out}'} 
  \bigg(\Delta \frac{ {\bf p}_{a_1}^2}{3 E_{a_1} } - \Delta' \frac{ {\bf p}_{a_1}^{\prime 2} }{3 E_{a_1}^\prime } \bigg)
\nonumber\\
&\times  \Big[  \big( f_{a_1'} \cdots f_{b_1'} \cdots (1 \pm f_{a_1})  \cdots (1\pm f_{b_1}) \cdots \big)
-f_{a_1} \cdots f_{b_1} \cdots    (1\pm f_{a_1'}) \cdots (1\pm f_{b_1'}) \cdots \Big] \,, \label{eq:generic-collision-1}
\end{align}
 where $g_{a_1}$ is the internal degrees of freedom of the particle $a_1$, 
 \begin{align}
 d\Pi_i \equiv \frac{d^3p_i}{(2\pi)^3 2E_i} \,,
 \end{align}
   $1\pm f_i$   with the plus or minus sign encodes the Bose enhancement or Pauli blocking, corresponding to
 the distribution
 \begin{align}
f_i (E_i, T_i) = \frac{1}{e^{(E_i -\mu_i)/T_i} \mp 1} \,,
\end{align}
with $\mu_i$ the chemical potential,
$|M|^2$ is the spin-summed amplitude squared,
  $S_{\rm in}$ and $S_{\rm out}'$ are respectively the symmetry factors for identical incoming and outgoing particles, $\Delta$ and $\Delta'$ are respectively the number
  of  $a_1$ in the initial and final states, and $S\equiv 1/2\, (1)$ for an elastic (inelastic) interaction. 
We consider the reactions mostly contributed by the phase space region $f_i \ll1$, i.e.,  $1\pm f_i \simeq 1$. Therefore the distributions can be well approximated as
\begin{align}
f_{i} =  e^{-(E_{i}-\mu_i)/T_i} (1\pm f_i) \simeq e^{-(E_{i}-\mu_i)/T_i}  \,.
\end{align}

The collision terms in the temperature Boltzmann equations are given by
\begin{align}
& g_X \int \frac{d^3p_X}{(2\pi)^3} \, C \Big[f_X\cdot \frac{{\bf p}_X^2}{3 E_X} \Big]
= - (2-\delta_H^X) n_X \gamma_X(T)  (T_X - T)
\nonumber\\
  & -   
   \bigg(   
             \langle \sigma v \cdot \frac{{\bf p}_X^2}{3 E_X} \rangle_{X X \to \sum_{ij} \text{SM}_i \text{SM}_j} (T_X) \,   n_X^2
            -   \langle \sigma v \cdot \frac{{\bf p}_X^2}{3 E_X} \rangle_{X X \to \sum_{ij} \text{SM}_i \text{SM}_j} (T) \,    
               \big( n_X^{\text{eq} } (T) \big)^2     \bigg)
 \nonumber\\
 &  +
     \frac{9m_S^2 -4m_X^2}{54 m_S}
         \langle \sigma v^2 \rangle_{SSS\to XX} 
                    \bigg(  n_S^3   - \frac{  \big(n_S^{\text{eq} } (T_X) \big)^3   n_X^2 } {(n_X^{\text{eq}} (T_X))^2 }  \bigg)  
\nonumber\\
 & + \frac{ m_S^2 (2m_X+3m_S)(2m_X + m_S)}{4 (m_X+ 2m_S) ( 2 m_X^2 +4 m_X m_S + 3 m_S^2)} 
             \langle \sigma v^2 \rangle_{XSS\to XS} 
                     \bigg( n_X n_S^2    -  n_S^{\text{eq}} (T_X)  n_X n_S    \bigg)  
\nonumber\\
 &
   + \frac{ (4 m_X^2 - m_S^2) (16 m_X^2 - m_S^2)}{ 108 m_X ( 10m_X^2 - m_S^2)}  
   \langle \sigma v^2 \rangle_{XXX\to XS} 
   \bigg( n_X^3   - \frac{ (n_X^{\text{eq}}  (T_X))^2  n_X n_S} {n_S^{\text{eq}}(T_X) }  \bigg) 
  \,, \label{eq:col-t-X}
\end{align}
and
\begin{align}
 &  g_S \int \frac{d^3p_S}{(2\pi)^3} \, C \Big[f_S\cdot \frac{{\bf p}_S^2}{3 E_S} \Big] 
=   - (2-\delta_H^S) n_S \gamma_S(T) (T_S - T)
\nonumber\\
  & -   
   \bigg(   
             \langle \sigma v \cdot \frac{{\bf p}_S^2}{3 E_S} \rangle_{S S \to \sum_{ij} \text{SM}_i \text{SM}_j} (T_S) \,   n_S^2
            -   \langle \sigma v \cdot \frac{{\bf p}_S^2}{3 E_S} \rangle_{S S \to \sum_{ij} \text{SM}_i \text{SM}_j} (T) \,    
               \big( n_S^{\text{eq} } (T) \big)^2     \bigg)
\nonumber\\
  & -   \bigg(
            \langle \Gamma_S \frac{{\bf p}_S^2}{3 E_S} \rangle_{S \to \sum_{ij} \text{SM}_i \text{SM}_j^{(*)}} (T_S) n_S
             -  \langle \Gamma_S \frac{{\bf p}_S^2}{3 E_S} \rangle_{S \to \sum_{ij} \text{SM}_i \text{SM}_j^{(*)}} (T) n_S^{\text{eq}} (T) 
                                          \bigg)                       
\nonumber\\
 &  +
     \frac{(4m_X^2-m_S^2)(16m_X^2 -m_S^2)}{108 m_X(8m_X^2+m_S^2)}
             \langle \sigma v^2 \rangle_{XXX\to XS} 
                    \bigg(  n_X^3   - \frac{  \big(n_X^{\text{eq} } (T_X) \big)^2   n_X n_S} {n_S^{\text{eq}} (T_X) }  \bigg)  
\nonumber\\
 & + \frac{(2m_X+3m_S)(2m_X-m_S)}{6 (2m_X+m_S)} 
             \langle \sigma v^2 \rangle_{XXS\to SS} 
                     \bigg( n_X^2 n_S    - \frac{ \big(n_X^{\text{eq}} (T_X) \big)^2 n_S^2  } {n_S^{\text{eq}} (T_X) }  \bigg)  
\nonumber\\
 &
   + \frac{ m_S(2m_X+m_S)(2m_X+3m_S)}{4(m_X+2m_S)(4m_X+5m_S)}  
   \langle \sigma v^2 \rangle_{XSS\to XS} 
   \bigg( n_X n_S^2   - n_S^{\text{eq}}  (T_X)  n_X n_S   \bigg) 
\nonumber\\
 & 
  + \frac{5}{54}  m_S \langle \sigma v^2 \rangle_{SSS\to SS} \bigg( n_S^3   - n_S^{\text{eq}}(T_X)  n_S^2   \bigg) 
  \,, \label{eq:col-t-S}
\end{align}
where
\begin{align}
n_i^{\rm eq} (T_j) = g_i  \frac{m_i^2 T_j} {2\pi^2} K_2(m_i/T_j)
\end{align}
is the equilibrium density of the $i (\equiv X$ or $S)$ particle at temperature $T_j (\equiv T, T_X$ or $T_S)$. Note that here it is unnecessary to consider the contributions arising from $X S\leftrightarrow X S$ and  $X X\leftrightarrow S S$ since these two effects are canceled by each other after adding the two collision terms given in Eqs.~(\ref{eq:col-t-X}) and (\ref{eq:col-t-S}).

 \subsection{The temperature weighted thermal average of the $S$ decay width}\label{app:thermal-temp-width}
 
 The temperature weighted thermal average of the $S$ decay width is relevant to the kinetic energy transfer via $S \leftrightarrow \sum_{ij} \text{SM}_i \text{SM}_j^{(*)}$ \footnote{ ${\rm SM}_j^*$ means an off-shell particle, and therefore $\text{SM}_i \text{SM}_j^{*}$ corresponds to a three-body decay mode. Here we consider dominant two- and three-body decay modes.}. The result is given by
\begin{align}
  \langle \Gamma_S \frac{{\bf p}_S^2}{3 E_S} \rangle_{S \to \sum_{ij} \text{SM}_i \text{SM}_j^{(*)}} (T_k)
 = \frac{ 2\pi^2  \Gamma_S }{ 3 m_S T_k  K_2(m_S/T_k)}  \int \frac{d^3 p_S}{(2\pi)^3} \frac{{\bf p}_S^2}{E_S^2} e^{-E_S/T_k}  \,,
\end{align}
and is simply denoted as
\begin{align}
  \langle \Gamma_S \frac{{\bf p}_S^2}{3 E_S} \rangle_{S \to \sum_{ij} \text{SM}_i \text{SM}_j^{(*)}} (T_k)
\equiv \langle \Gamma_S \rangle_{T_k} \delta_\Gamma (T_k)  T_k 
\label{app:deltag}
\end{align}  
in the text, where
 \begin{equation}
 \langle \Gamma_S \rangle_{T_k}  = \Gamma_S \frac{K_1(m_S/T_k)}{K_2(m_S/T_k)} 
\end{equation}
is the thermal average of the decay width of $S$, and
$T_k \equiv T_S$ or $T$.

 \subsection{The temperature weighted thermal averages of annihilation processes, $XX \to {\rm SM~SM}$ and $SS \to {\rm SM~SM}$}\label{app:thermal-temp-ann}
 
The temperature-weighted thermal averages of $2 \to 2$  annihilations, $XX \to {\rm SM~SM}$ and $SS \to {\rm SM~SM}$, are relevant to the kinetic energy transfer between the hidden sector and SM bath. We use the result \cite{Yang:2019bvg},
\begin{align}
 \langle \sigma v  \cdot \frac{{\bf p}_i^2}{3 E_i} \rangle (T_k)
  \simeq  & \frac{1}{48 m_i^4  K_2^2 (m_i/T_k)}       
  \int_{4m_i^2}^{\infty} ds \,  \theta(s-4m_{\rm SM}^2)  (\sigma v_{\text{lab}} )  \sqrt{s-4m_i^2} 
  \nonumber\\
      & \times 
    \bigg[ ( s+ 2 m_i^2) K_1 \left(\frac{\sqrt{s}}{T_k} \right)
         +  \left( \frac{s-4m_i^2}{2} \frac{\sqrt{s}}{T_k} + \frac{ 4T_k ( s+ 2 m_i^2)}{\sqrt{s}}  \right) K_2 \left(\frac{\sqrt{s}}{T_k} \right)
    \bigg], 
\label{eq:thermal-temp-ann}
\end{align}
where the subscript $``i" \equiv X$ or $S$, and $T_k \equiv T_S$ or $T$.

 \subsection{The temperature weighted thermal averages of  $3 \to 2$  annihilation rates}\label{app:thermal-temp-3to2} 

In the calculation of the temperature weighted thermal averages of $3 \to 2$  annihilations, $ \langle \sigma v^2  \cdot \frac{{\bf p}_i^2}{3 E_i} \rangle_{3\to 2}$ with  the subscript $``i" \equiv X$ or $S$,
we use the non-relativistic limit $E_{\rm in} \approx m_{\rm in}$ for an incoming particles, where $\rm ``in" \equiv$  ``incoming particle". Taking $ \langle \sigma v^2  \cdot \frac{{\bf p}_X^2}{3 E_X} \rangle_{SSS\to XX}$ as an example, we have
\begin{align}
   \langle \sigma v^2  \cdot \frac{{\bf p}_X^2}{3 E_X} \rangle_{SSS\to XX}
\approx
   \langle \sigma v^2 \rangle_{SSS\to XX}  \cdot 2 \bigg(\frac{{\bf p}_X^2}{3 E_X} \bigg)_{SSS\to XX} \,,
   \label{eq:thermal-3to2-example}
\end{align}
where
\begin{align}
\bigg(\frac{{\bf p}_X^2}{3 E_X} \bigg)_{SSS\to XX}  
& = \frac{1}{3} \frac{E_X^2 - m_X^2}{ E_X}
      \approx \frac{1}{3} \frac{ (3m_S/2)^2 - m_X^2}{ 3m_S/2 } \nonumber\\
&\approx \frac{1}{9} \frac{ 9 m_S^2 - 4 m_X^2}{ m_S }
 \,.
\end{align}
The right-hand side of Eq.~(\ref{eq:thermal-3to2-example}) contains a factor of 2 
because two $X$ particles are produced in the interaction. Therefore, the  contribution of the $ SSS\leftrightarrow XX$ interaction  to the DM temperature Boltzmann equations given in Eq.~(\ref{eq:col-t-X}) is in the form
\begin{equation}
     \frac{9m_S^2 -4m_X^2}{3! \cdot 9 m_S}
         \langle \sigma v^2 \rangle_{SSS\to XX} 
                    \bigg(  n_S^3   - \frac{  \big(n_S^{\text{eq} } (T_X) \big)^3   n_X^2 } {(n_X^{\text{eq}} (T_X))^2 }  \bigg)  \,,
\end{equation}
where the factor of $1/3!$ accounts for the identical incoming particles, while the symmetry factor for the identical outgoing particles is included in $ \langle \sigma v^2 \rangle_{SSS\to XX}$ by definition \cite{Yang:2019bvg}.

 \subsection{The temperature weighted thermal-average of elastic scatterings between the hidden sector and SM bath}\label{app:thermal-temp-elastic}

The thermal-average kinetic energy transfer between the hidden species  $i (\equiv X$ or $S)$ and relativistic SM particle $f$  via the elastic scattering  $i~f \leftrightarrow i~f$  can be described by
a semi-relativistic Fokker-Planck-type equation \cite{Yang:2019bvg,Bringmann:2006mu,Bringmann:2009vf,Gondolo:2012vh,Visinelli:2015eka,Binder:2016pnr,Binder:2017rgn},
\begin{align}
 g_i  \int \frac{d^3 p_i}{(2\pi)^3}  \,  C\Big[f_i\cdot \frac{{\bf p}_i^2}{3 E_i}\Big]_{i\, f \leftrightarrow i\, f}
 & \simeq 
  \gamma_i^f (T) \,  g_i \int   \frac{d^3p_i}{(2\pi)^3 }   \frac{ {\bf p}_i^2}{3 E_i} \frac{\partial}{ \partial {\bf p}_i} \cdot 
        \left( {\bf p}_i f_i (T_i) + E_i T \frac{\partial f_i (T_i)}{\partial{\bf p}_i} \right) \nonumber\\
& \simeq - (2-\delta_{H}^i (T_i) ) n_i\  \gamma_i^f (T)   \,  (T_i -T) \,.
\end{align}
The momentum transfer rate $\gamma_i^f$  reads
\begin{align}
\gamma_i^f (T) =
\frac{1}{g_i}  \frac{1}{6 m_i T} \int  \frac{d^3 k}{(2\pi)^3} f_f (T) (1 \mp f_f (T)) \frac{|{\bf k}|}{\sqrt{{\bf k}^2 +m_f^2}} 
\int_{-4 {\bf k}^2}^0 dt (-t) \frac{d \sigma_{i f \to i f }}{dt} \,, \label{eq:gammah}
\end{align} 
where the term $1\mp f_f$ with minus (plus) sign stands for the SM $f \equiv$ fermion (boson), and the differential elastic scattering cross section is
\begin{align}
\frac{d \sigma_{i f \to i f}}{dt} = \frac{1}{64\pi m_i^2 {\bf k}^2}  |M_{i f \to i f }|^2 \,.
\end{align}
Thus the total momentum transfer rate is given by $\gamma_i = \sum_f \gamma_i^f$.
Here, $|M_{i f\to i f}|^2$ is the squared amplitude, up to the order of $s_\alpha^2$ and summed over initial and final internal DoFs. We obtain, for  $f \equiv$ fermion,
\begin{align}
  & |M_{X f \to X f }|^2 =8  N_c^f  \,
     \Big( 2 + \frac{ (2m_X^2-t)^2}{4 m_X^4} \Big)
   (4 m_f^2 -t )  g_{\rm dm}^2 m_X^2
   \bigg( \frac{c_\alpha \, g_{S ff}}{t- m_S^2} +  \frac{s_\alpha \, g_{h ff}}{t- m_h^2} \bigg)^2  
  \,,  \\
  & |M_{S f \to S f }|^2 =  2 N_c^f  (4 m_f^2 -t )  \bigg( \frac{g_{SSS}\, g_{S ff}}{t- m_S^2} +  \frac{g_{SSh} \, g_{h ff}}{t- m_h^2} \bigg)^2
  \,,
\end{align}
with $N_c^{f} \equiv 3\, (1)$ for quarks (leptons), and for  $f \equiv$ bosons,
\begin{align}
  & |M_{X V\to X V}|^2 =4 \,
     \Big( 2 + \frac{ (2m_X^2-t)^2}{4 m_X^4} \Big)     \Big( 2 + \frac{ (2m_V^2-t)^2}{4 m_V^4} \Big)
   g_{\rm dm}^2 m_X^2
   \bigg( \frac{c_\alpha \, g_{S VV}}{t- m_S^2} +  \frac{s_\alpha \, g_{h VV}}{t- m_h^2} \bigg)^2  \,, \\
    & |M_{S V \to S V}|^2 =    \Big( 2 + \frac{ (2m_V^2-t)^2}{4 m_V^4} \Big) \bigg( \frac{g_{SSS}\, g_{S VV}}{t- m_S^2} +  \frac{g_{SSh} \, g_{h VV}}{t- m_h^2} \bigg)^2
  \,, \hskip0.5cm  \text{with~} V\equiv W, Z \,,\\
  & |M_{X h\to X h}|^2 =4 \,
     \Big( 2 + \frac{ (2m_X^2-t)^2}{4 m_X^4} \Big)     g_{\rm dm}^2 m_X^2
   \bigg( \frac{c_\alpha \, g_{Shh}}{t- m_S^2} +  \frac{s_\alpha \, g_{h hh}}{t- m_h^2} \bigg)^2  \,, \\
   & |M_{S h \to S h }|^2 =  \bigg( \frac{g_{SSS}\, g_{S hh}}{t- m_S^2} +  \frac{g_{SSh} \, g_{h hh}}{t- m_h^2} \bigg)^2
  \,, \\
   & |M_{X g\to X g}|^2 =64 \,  \Big( 2 + \frac{ (2m_X^2-t)^2}{4 m_X^4} \Big) 
   g_{\rm dm}^2 m_X^2 \, t^2
   \bigg( \frac{c_\alpha \, g_{S gg}(\sqrt{-t}) }{t- m_S^2} +  \frac{s_\alpha \, g_{h gg}(\sqrt{-t}) }{t- m_h^2} \bigg)^2  \,, \\
    & |M_{S g\to S g}|^2 =16  \, t^2
   \bigg( \frac{g_{SSS} \, g_{S gg}(\sqrt{-t} )}{t- m_S^2} +  \frac{g_{SSh} \, g_{h gg} (\sqrt{-t}) }{t- m_h^2} \bigg)^2  \,, 
    \end{align}
We assume that $X g \to X g$ and $S g \to S g$ are active only when $T>T_{\rm QCD}$.  Moreover, we use the QCD renormalization scale $\mu=T/1.76$ for $T\geq1$~GeV and $\mu =0.57$~GeV for $ T_{\rm QCD} \leq T < 1$~GeV, consistent with the scale choice in Appendix \ref{app:hidden2SM}  by replacing the typical scale $\sqrt{s}$ with the bath temperature $T$, where we adopt  the QCD phase transition temperature $T_{\rm QCD}=150$~MeV.

\section{Boltzmann equations}\label{app:boltz}

One can then deduce the time derivative in terms of the bath temperature,
\begin{align}
\frac{d}{dt} 
      =  \frac{\tilde{H}}{x} 
   \frac{h_{\rm eff}(T)}{\tilde{h}_{\rm eff}(T)}  \frac{d}{dx}\,,
\end{align}
where $x = m_X / T$ is the dimensionless temperature variable, $\tilde{h}_{\rm eff}  \equiv h_{\rm eff} [1+(1/3) (d\ln h_{\rm eff} / d\ln T)]$,  $H^2  = (8\pi /3) G_N \rho$ is the Hubble parameter with $\rho$ the total energy density of the Universe, and $\tilde{H} \equiv x^2 H$.
Here 
\begin{align}
h_{\rm eff}(T) =\bigg( \frac{2\pi^2}{45} T^3 \bigg)^{-1} s(T) \,,
\end{align}
with $s$ being the total entropy density of the Universe.

Besides introducing dimensionless quantities,
\begin{align}
&Y_X (T_X) \equiv \frac{n_X(T_X)}{s (T)}, \qquad  Y_S (T_S)\equiv \frac{n_S(T_S)}{s(T)} \,,    \qquad {\text{(yields)}}  \nonumber\\
& y= \frac{T_X}{T}\,,    \qquad {\text{(temperature ratio)}} 
\label{app:dim-less-variables}
\end{align}
as functions of a dimensionless variable $x$ and letting $T_X=T_S$, we obtain the coupled Boltzmann equations of the yields,
\begin{align}
\frac{d Y_X}{dx} =
   &
  \frac{\tilde{h}_{\rm eff}(T)}{h_{\rm eff}(T)}   \frac{1}{xH}  \nonumber\\
   &\times \Bigg[ 
       s \langle \sigma v\rangle_{SS\to XX} (T_X)  \bigg(  Y_S^2 
         -   \frac{(Y_S^{\text{eq}} (T_X))^2 }{ (Y_X^{\text{eq}} (T_X)^2}   Y_X^2  \bigg)  \nonumber\\
     & \ 
        + \frac{s^2}{3} \langle \sigma v^2 \rangle_{SSS\to XX} 
                 \bigg( Y_S^3 - \frac{(Y_S^{\text{eq}} (T_X))^3 }{(Y_X^{\text{eq}} (T_X))^2}  Y_X^2  \bigg)  
           - s^2 \langle \sigma v^2 \rangle_{XXS\to SS}
                  \bigg( Y_X^2 Y_S -    \frac{(Y_X^{\text{eq}} (T_X))^2}{Y_S^{\text{eq}} (T_X)}  Y_S^2   \bigg)    
            \nonumber\\
     &  \   
         - \frac{s^2}{3} \langle \sigma v^2 \rangle_{XXX\to XS}               
          \bigg( Y_X^3 -\frac{  (Y_X^{\text{eq}}(T_X))^2 }{ Y_S^{\text{eq}}(T_X) }   Y_X   Y_S \bigg)      
             \nonumber\\
     & \ 
         -   s \Big(  \langle \sigma v \rangle_{XX\to \sum_i {\rm SM}_i {\rm SM}_i}  (T_X) Y_X^2 
            - \langle \sigma v \rangle_{XX\to \sum_i {\rm SM}_i {\rm SM}_i} (T) \, ( Y_X^{\text{eq}} (T) )^2 \Big)
        \Bigg] \,,    
  \label{eq:boltz-YX}
 \end{align}
\begin{align}
\frac{d Y_S}{dx} =
 & 
   \frac{\tilde{h}_{\rm eff}(T)}{h_{\rm eff}(T)}   \frac{1}{xH}  \nonumber\\
 &\times \Bigg[ 
    - s \langle \sigma v\rangle_{SS\to XX} (T_X)    \bigg(   Y_S^2 
         -   \frac{(Y_S^{\text{eq}} (T_X))^2 }{ (Y_X^{\text{eq}} (T_X)^2}   Y_X^2  \bigg)  \nonumber\\
 &  -  \Gamma_{S}    \bigg( \frac{K_1(m_S/T_X)}{K_2( m_S/T_X)} Y_S
         - \frac{K_1(m_S/T)}{K_2( m_S/T)} Y_S^{\text{eq}}(T) \bigg)  
   \nonumber\\
 &  -    s \Big(  \langle \sigma v \rangle_{SS\to \sum_i {\rm SM}_i {\rm SM}_i}(T_X)  Y_S^2 
         - \langle \sigma v \rangle_{SS\to \sum_i {\rm SM}_i {\rm SM}_i} (T) \, ( Y_S^{\text{eq}} (T) )^2 \Big)
   \nonumber\\
 &  
         +  
              \frac{s^2}{6}  \langle \sigma v^2 \rangle_{XXX\to XS}
                      \bigg( Y_X^3 -  Y_X  Y_S    \frac{ (Y_X^{\text{eq}}(T_X))^2 }{ Y_S^{\text{eq}}(T_X)}   \bigg)   
             - \frac{s^2}{6}    \langle \sigma v^2 \rangle_{SSS\to SS}  
              \Big( Y_S^3 - Y_S^2  Y_S^{\text{eq}} (T_X) \Big) 
     \nonumber\\
 &                
      +    \frac{s^2}{2}  \langle \sigma v^2 \rangle_{XXS\to SS}
                   \bigg( Y_X^2 Y_S -   \frac{(Y_X^{\text{eq}} (T_X))^2 }{Y_S^{\text{eq}}(T_X)} Y_S^2 \bigg)    
  -    \frac{s^2}{2}  \langle \sigma v^2 \rangle_{SSS\to XX}
               \bigg( Y_S^3 - \frac{ (Y_S^{\text{eq}}(T_X))^3}{(Y_X^{\text{eq}}(T_X))^2} Y_X^2 \bigg)
     \nonumber\\
 &   -  \frac{s^2}{2}  \langle \sigma v^2 \rangle_{XSS\to XS}
              \Big( Y_X Y_S^2 - Y_X Y_S  Y_S^{\text{eq}}(T_X)  \Big) 
\Bigg]
    \,,  \label{eq:boltz-YS}
 \end{align}
and the evolution of the temperature ratio, rewritten from Eq.~(\ref{eq:boltz-t-hidden}),
\begin{align}
\frac{dy}{dx} =
& \frac{y}{x}
 - \frac{y}{Y_X + Y_{S}}
\bigg( \frac{dY_X}{dx} +  \frac{dY_S}{dx} 
\bigg)
\nonumber \\
&  + 
\frac{1}{Y_X + Y_{S}}
 \frac{\tilde{h}_{\rm eff}(T)}{h_{\rm eff}(T)}
\Bigg\{
 -  (2 - \delta_{H}^X)  
  \Big(\frac{y}{x} +  \frac{\gamma_X}{x H}  (y-1) 
  \Big) Y_X
  \nonumber\\
 & \  \
  - \frac{s}{ x H } 
     \Big(y\,  Y_X^2   \widetilde{\langle \sigma v \rangle}_{XX\to \sum_{ij} {\rm SM}_i {\rm SM}_j} (T_X)
     -  (Y_X^{\text{eq}} (T) )^2   \widetilde{\langle \sigma v \rangle}_{ XX\to \sum_{ij} {\rm SM}_i {\rm SM}_j } (T)   \Big)
   \nonumber \\
&  \  \
  +  \frac{s^2}{m_X H}   \Bigg[
    \frac{(4m_X^2-m_S^2)(16m_X^2 -m_S^2)}{108 m_X(10m_X^2 -m_S^2)}
     \langle \sigma v^2 \rangle_{XXX\to XS}
    \bigg( Y_X^3- \frac{  Y_X Y_S  \big( Y_X^{\text{eq}} (T_X) \big)^2}{ Y_S^{\text{eq}}(T_X) } \bigg)   
    \nonumber\\
& \  \  \  \
  + \frac{ m_S^2 (2m_X+m_S)(2m_X+3m_S)}{4(m_X+2m_S)(2m_X^2 +4m_X m_S + 3m_S^2)}  
    \langle \sigma v^2 \rangle_{XSS\to XS}
    \Big( Y_X Y_S^2 - Y_X Y_S Y_S^{\text{eq}}(T_X)  \Big) 
 \nonumber\\
& \  \  \  \
  + \frac{(2m_X +3m_S)(3m_S -2m_X)}{54 m_S}
     \langle \sigma v^2 \rangle_{SSS\to XX}
    \bigg( Y_S^3- \frac{  Y_X^2  \big( Y_S^{\text{eq}} (T_X) \big)^3}{ \big( Y_X^{\text{eq}}(T_X)  \big)^2 } \bigg)   \Bigg]
    \nonumber\\
& \  \
 -  (2 - \delta_{H}^S)  
   \Big(\frac{y}{x} +  \frac{\gamma_S}{x H}  (y-1) 
   \Big) Y_S
   \nonumber\\
&  \  \
 - \frac{s}{ x H } 
    \Big(y\,  Y_S^2   \widetilde{\langle \sigma v \rangle}_{SS\to \sum_{ij} {\rm SM}_i {\rm SM}_j} (T_X)
     -  (Y_S^{\text{eq}} (T) )^2   \widetilde{\langle \sigma v \rangle}_{ SS\to \sum_{ij} {\rm SM}_i {\rm SM}_j } (T)   \Big)
   \nonumber \\
&  \  \ 
    - \frac{ \Gamma_{S}}{x H}  
        \bigg(    \frac{K_1(m_S/T_X)}{K_2(m_S/T_X)} Y_S \delta_\Gamma (T_X)  y
    -\frac{K_1(m_S/T)}{K_2(m_S/T)}  Y_S^{\text{eq}} (T)  \delta_\Gamma (T) 
         \bigg)
   \nonumber\\
 &  \   \
   +  \frac{s^2}{m_X H}   \Bigg[
     \frac{(4m_X^2-m_S^2)(16m_X^2 -m_S^2)}{108 m_X(8m_X^2+m_S^2)}
     \langle \sigma v^2 \rangle_{XXX\to XS}
     \bigg( Y_X^3- \frac{  Y_X Y_S  \big( Y_X^{\text{eq}} (T_X) \big)^2}{ Y_S^{\text{eq}}(T_X) } \bigg)   
    \nonumber\\
&\  \  \  \
 +  \frac{(2m_X+3m_S)(2m_X-m_S)}{6 (2m_X+m_S)} 
    \langle \sigma v^2 \rangle_{XXS\to SS}
    \bigg( Y_X^2 Y_S - \frac{Y_S^2 \big( Y_X^{\text{eq}}(T_X) \big)^2}{Y_S^{\text{eq}} (T_X) }  \bigg)   
 \nonumber\\
& \  \  \  \
+ \frac{ m_S(2m_X+m_S)(2m_X+3m_S)}{4(m_X+2m_S)(4m_X+5m_S)}  
    \langle \sigma v^2 \rangle_{XSS\to XS}
   \Big( Y_X Y_S^2 - Y_X Y_S Y_S^{\text{eq}}(T_X)  \Big) 
 \nonumber\\
& \   \  \  \
+ \frac{5}{54}  m_S 
 \langle \sigma v^2 \rangle_{SSS\to SS}
   \Big( Y_S^3 - Y_S^2  Y_S^{\text{eq}}(T_X)  \Big)
   \Bigg] 
   \Bigg\}
 \,, 
\end{align}
where we have set $T_X=T_S$, and $\widetilde{\langle \sigma v \rangle}_{SS\to \sum_{ij} {\rm SM}_i {\rm SM}_j} (T_k) 
\equiv  \langle \sigma v \cdot \frac{{\bf p}_S^2}{3 E_S} \rangle_{S S \to \sum_{ij} \text{SM}_i \text{SM}_j} (T_k) /T_k$ , with $T_k =T_S$ or $T$.

\section{The elastic decoupling of $XS \leftrightarrow XS$}\label{app:x-ed}

Considering the Boltzmann equation of the DM temperature alone (see also Eqs.~(\ref{eq:boltz-t-1}) and (\ref{eq:col-t-X})), and further including the $X~S\leftrightarrow X~S$ term, we have
\begin{align}
  \frac{d T_X}{dt}  + (2 - \delta_H^X) H T_X 
& =    -(2 - \delta_H^X) \gamma_{XS} (T_X - T_S) 
   +  \cdots \,,
 \label{app:boltz-t-X}
\end{align}
where momentum relaxation rate $\gamma_{XS}$ for $X~S\leftrightarrow X~S$ is given by \cite{Yang:2019bvg}
\begin{align}
\gamma_{XS} 
& =
 \frac{1}{ 768 \pi^3 m_X^3 g_X T_S} \int _{m_S}^\infty d E_S \,   f_S (T_S)  \, (1 + f_S (T_S))
\int_{-4 {\bf p}_S^2}^0 dt (-t)  |M_{X S \to X S }|^2  \nonumber\\
& \simeq 
  \frac{1}{g_X} \mathop{\hspace{-15ex}  |M_{X S \to X S }|^2_{t=0} }_{\hspace{10.5ex}s=m_X^2+2m_X m_S +m_S^2}
     \frac{1}{12 \pi^3}  \frac{m_S^2}{m_X} \frac{T_S^2}{m_X^2} 
  \bigg( 1 + 3 \frac{T_S}{m_S} + 3 \frac{T_S^2}{m_S^2} \bigg)  e^{-m_S/T_S}\,,  
  \label{app:gammaX}
\end{align}
with
\begin{align}
   \mathop{\hspace{-15ex}  |M_{X S \to X S }|^2_{t=0} }_{\hspace{10.5ex}s=m_X^2+2m_X m_S +m_S^2} 
 \simeq 4 g_X g_{\rm dm}^2 \cos^2\alpha 
   \frac{ \big( g_{\rm dm}  \cos\alpha  \, m_S^4 - g_{SSS} m_X (4m_X^2 -m_S^2) \big)^2}{ m_S^4 (4m_X^2 -m_S^2)^2}
    \,.
\end{align}
We study the case that, resulting from a large ${\rm SM~SM} \to S$ rate,  $T_S=T$ can be maintained until after the time of $T_X \not= T$. As such, the elastic decoupling of $XS \leftrightarrow XS$ can be set by $H \simeq \gamma_{XS}$.


\begin{thebibliography}{99}


%\cite{Adam:2015rua}
\bibitem{Adam:2015rua} 
  R.~Adam {\it et al.} [Planck Collaboration],
  %``Planck 2015 results. I. Overview of products and scientific results,''
  Astron.\ Astrophys.\  {\bf 594}, A1 (2016)
%  doi:10.1051/0004-6361/201527101
  [arXiv:1502.01582 [astro-ph.CO]].
  %%CITATION = doi:10.1051/0004-6361/201527101;%%
  %831 citations counted in INSPIRE as of 20 Nov 2019
  
 %\cite{Ade:2015xua}
\bibitem{Ade:2015xua} 
  P.~A.~R.~Ade {\it et al.} [Planck Collaboration],
  %``Planck 2015 results. XIII. Cosmological parameters,''
  Astron.\ Astrophys.\  {\bf 594}, A13 (2016)
%  doi:10.1051/0004-6361/201525830
  [arXiv:1502.01589 [astro-ph.CO]].
  %%CITATION = doi:10.1051/0004-6361/201525830;%%
  %7941 citations counted in INSPIRE as of 20 Nov 2019
  
%\cite{Aprile:2017iyp}
%\bibitem{Aprile:2017iyp} 
%  E.~Aprile {\it et al.} [XENON Collaboration],
%  ``First Dark Matter Search Results from the XENON1T Experiment,''
%  Phys.\ Rev.\ Lett.\  {\bf 119}, no. 18, 181301 (2017);
 % doi:10.1103/PhysRevLett.119.181301
%  [arXiv:1705.06655 [astro-ph.CO]];
  %%CITATION = doi:10.1103/PhysRevLett.119.181301;%%
  %560 citations counted in INSPIRE as of 14 Feb 2019

%\cite{XENON:2018voc}
\bibitem{XENON:2018voc}
E.~Aprile \textit{et al.} [XENON],
%``Dark Matter Search Results from a One Ton-Year Exposure of XENON1T,''
Phys. Rev. Lett. \textbf{121} (2018) no.11, 111302
%doi:10.1103/PhysRevLett.121.111302
[arXiv:1805.12562 [astro-ph.CO]].
%1619 citations counted in INSPIRE as of 30 May 2022

%\cite{PandaX-4T:2021bab}
\bibitem{PandaX-4T:2021bab}
Y.~Meng \textit{et al.} [PandaX-4T],
%``Dark Matter Search Results from the PandaX-4T Commissioning Run,''
Phys. Rev. Lett. \textbf{127} (2021) no.26, 261802
%doi:10.1103/PhysRevLett.127.261802
[arXiv:2107.13438 [hep-ex]].
%107 citations counted in INSPIRE as of 30 May 2022

%\cite{XENON:2019rxp}
\bibitem{XENON:2019rxp}
E.~Aprile \textit{et al.} [XENON],
%``Constraining the spin-dependent WIMP-nucleon cross sections with XENON1T,''
Phys. Rev. Lett. \textbf{122} (2019) no.14, 141301
%doi:10.1103/PhysRevLett.122.141301
[arXiv:1902.03234 [astro-ph.CO]].
%197 citations counted in INSPIRE as of 30 May 2022

%\cite{Griest:1990kh}
\bibitem{Griest:1990kh}
K.~Griest and D.~Seckel,
%``Three exceptions in the calculation of relic abundances,''
Phys. Rev. D \textbf{43} (1991), 3191-3203
% doi:10.1103/PhysRevD.43.3191
%1373 citations counted in INSPIRE as of 26 May 2022

%\cite{Tulin:2012uq}
\bibitem{Tulin:2012uq}
S.~Tulin, H.~B.~Yu and K.~M.~Zurek,
%``Three Exceptions for Thermal Dark Matter with Enhanced Annihilation to $\gamma \gamma$,''
Phys. Rev. D \textbf{87} (2013) no.3, 036011
% doi:10.1103/PhysRevD.87.036011
[arXiv:1208.0009 [hep-ph]].
%70 citations counted in INSPIRE as of 31 May 2022


%\cite{Jackson:2013pjq}
\bibitem{Jackson:2013pjq}
C.~B.~Jackson, G.~Servant, G.~Shaughnessy, T.~M.~P.~Tait and M.~Taoso,
%``Gamma-ray lines and One-Loop Continuum from s-channel Dark Matter Annihilations,''
JCAP \textbf{07} (2013), 021
% doi:10.1088/1475-7516/2013/07/021
[arXiv:1302.1802 [hep-ph]].
%43 citations counted in INSPIRE as of 26 May 2022


%\cite{Jackson:2013tca}
\bibitem{Jackson:2013tca}
C.~B.~Jackson, G.~Servant, G.~Shaughnessy, T.~M.~P.~Tait and M.~Taoso,
%``Gamma Rays from Top-Mediated Dark Matter Annihilations,''
JCAP \textbf{07} (2013), 006
% doi:10.1088/1475-7516/2013/07/006
[arXiv:1303.4717 [hep-ph]].
%21 citations counted in INSPIRE as of 31 May 2022

%\cite{Delgado:2016umt}
\bibitem{Delgado:2016umt}
A.~Delgado, A.~Martin and N.~Raj,
%``Forbidden Dark Matter at the Weak Scale via the Top Portal,''
Phys. Rev. D \textbf{95}, no.3, 035002 (2017)
% doi:10.1103/PhysRevD.95.035002
[arXiv:1608.05345 [hep-ph]].
%14 citations counted in INSPIRE as of 04 Sep 2022

%\cite{DAgnolo:2015ujb}
\bibitem{DAgnolo:2015ujb}
R.~T.~D'Agnolo and J.~T.~Ruderman,
%``Light Dark Matter from Forbidden Channels,''
Phys. Rev. Lett. \textbf{115} (2015) no.6, 061301
% doi:10.1103/PhysRevLett.115.061301
[arXiv:1505.07107 [hep-ph]].
%136 citations counted in INSPIRE as of 24 May 2022


%\cite{DAgnolo:2020mpt}
\bibitem{DAgnolo:2020mpt}
R.~T.~D'Agnolo, D.~Liu, J.~T.~Ruderman and P.~J.~Wang,
%``Forbidden dark matter annihilations into Standard Model particles,''
JHEP \textbf{06} (2021), 103
% doi:10.1007/JHEP06(2021)103
[arXiv:2012.11766 [hep-ph]].
%5 citations counted in INSPIRE as of 24 May 2022

%\cite{Wojcik:2021xki}
\bibitem{Wojcik:2021xki}
G.~N.~Wojcik and T.~G.~Rizzo,
%``Forbidden scalar dark matter and dark Higgses,''
JHEP \textbf{04}, 033 (2022)
% doi:10.1007/JHEP04(2022)033
[arXiv:2109.07369 [hep-ph]].
%2 citations counted in INSPIRE as of 04 Sep 2022


%\cite{Yang:2019bvg}
\bibitem{Yang:2019bvg}
K.~C.~Yang,
%``Thermodynamic Evolution of Secluded Vector Dark Matter: Conventional WIMPs and Nonconventional WIMPs,''
JHEP \textbf{11} (2019), 048
% doi:10.1007/JHEP11(2019)048
[arXiv:1905.09582 [hep-ph]].
%9 citations counted in INSPIRE as of 04 Jun 2022

%\cite{Yang:2020vxl}
\bibitem{Yang:2020vxl}
K.~C.~Yang,
%``A potentially detectable gamma-ray line in the Fermi Galactic center excess \textemdash{} in light of one-step cascade annihilations of secluded (vector) dark matter via the Higgs portal,''
JHEP \textbf{07}, 148 (2020)
% doi:10.1007/JHEP07(2020)148
[arXiv:2001.04946 [hep-ph]].
%4 citations counted in INSPIRE as of 18 Aug 2022

%\cite{Cerdeno:2011tf}
\bibitem{Cerdeno:2011tf}
D.~G.~Cerdeno, T.~Delahaye and J.~Lavalle,
%``Cosmic-ray antiproton constraints on light singlino-like dark matter candidates,''
Nucl. Phys. B \textbf{854}, 738-779 (2012)
% doi:10.1016/j.nuclphysb.2011.09.020
[arXiv:1108.1128 [hep-ph]].
%26 citations counted in INSPIRE as of 25 Jul 2022



%\cite{Gondolo:1990dk}
\bibitem{Gondolo:1990dk} 
  P.~Gondolo and G.~Gelmini,
%  ``Cosmic abundances of stable particles: Improved analysis,''
  Nucl.\ Phys.\ B {\bf 360}, 145 (1991).
%  doi:10.1016/0550-3213(91)90438-4
  %%CITATION = doi:10.1016/0550-3213(91)90438-4;%%
  %660 citations counted in INSPIRE as of 28 Feb 2016

%\cite{pdg2022}
\bibitem{pdg2022}
R.L. Workman et al. (Particle Data Group), Prog. Theor. Exp. Phys. 2022, 083C01 (2022).
 
 %\cite{Baek:2012se}
\bibitem{Baek:2012se}
S.~Baek, P.~Ko, W.~I.~Park and E.~Senaha,
%``Higgs Portal Vector Dark Matter : Revisited,''
JHEP \textbf{05}, 036 (2013)
% doi:10.1007/JHEP05(2013)036
[arXiv:1212.2131 [hep-ph]], and references therein.
%166 citations counted in INSPIRE as of 08 Sep 2022
 
%\cite{Hall:2009bx}
\bibitem{Hall:2009bx}
L.~J.~Hall, K.~Jedamzik, J.~March-Russell and S.~M.~West,
%``Freeze-In Production of FIMP Dark Matter,''
JHEP \textbf{03}, 080 (2010)
% doi:10.1007/JHEP03(2010)080
[arXiv:0911.1120 [hep-ph]].
%947 citations counted in INSPIRE as of 21 Sep 2022

%\cite{Farina:2016llk}
\bibitem{Farina:2016llk}
M.~Farina, D.~Pappadopulo, J.~T.~Ruderman and G.~Trevisan,
%``Phases of Cannibal Dark Matter,''
JHEP \textbf{12}, 039 (2016)
%doi:10.1007/JHEP12(2016)039
[arXiv:1607.03108 [hep-ph]].
%86 citations counted in INSPIRE as of 20 Oct 2022

%\cite{Alekhin:2015byh}
\bibitem{Alekhin:2015byh}
S.~Alekhin, W.~Altmannshofer, T.~Asaka, B.~Batell, F.~Bezrukov, K.~Bondarenko, A.~Boyarsky, K.~Y.~Choi, C.~Corral and N.~Craig, \textit{et al.}
%``A facility to Search for Hidden Particles at the CERN SPS: the SHiP physics case,''
Rept. Prog. Phys. \textbf{79}, no.12, 124201 (2016)
% doi:10.1088/0034-4885/79/12/124201
[arXiv:1504.04855 [hep-ph]].
%738 citations counted in INSPIRE as of 25 Aug 2022

%\cite{SHiP:2018yqc}
\bibitem{SHiP:2018yqc}
C.~Ahdida \textit{et al.} [SHiP],
%``The experimental facility for the Search for Hidden Particles at the CERN SPS,''
JINST \textbf{14}, no.03, P03025 (2019)
% doi:10.1088/1748-0221/14/03/P03025
[arXiv:1810.06880 [physics.ins-det]].
%48 citations counted in INSPIRE as of 25 Aug 2022

%\cite{NA62:2017rwk}
\bibitem{NA62:2017rwk}
E.~Cortina Gil \textit{et al.} [NA62],
%``The Beam and detector of the NA62 experiment at CERN,''
JINST \textbf{12}, no.05, P05025 (2017)
% doi:10.1088/1748-0221/12/05/P05025
[arXiv:1703.08501 [physics.ins-det]].
%270 citations counted in INSPIRE as of 25 Aug 2022

%\cite{Akerib:2018lyp}
\bibitem{Akerib:2018lyp} 
  D.~S.~Akerib {\it et al.} [LUX-ZEPLIN Collaboration],
%  ``Projected WIMP Sensitivity of the LUX-ZEPLIN (LZ) Dark Matter Experiment,''
  arXiv:1802.06039 [astro-ph.IM].
  %%CITATION = ARXIV:1802.06039;%%
  %50 citations counted in INSPIRE as of 17 Feb 2019



%\cite{L3:1996ome}
\bibitem{L3:1996ome}
M.~Acciarri \textit{et al.} [L3],
%``Search for neutral Higgs boson production through the process e+ e- --\ensuremath{>} Z* H0,''
Phys. Lett. B \textbf{385}, 454-470 (1996)
% doi:10.1016/0370-2693(96)00987-2
%91 citations counted in INSPIRE as of 23 Aug 2022

%\cite{Winkler:2018qyg}
\bibitem{Winkler:2018qyg}
M.~W.~Winkler,
%``Decay and detection of a light scalar boson mixing with the Higgs boson,''
Phys. Rev. D \textbf{99}, no.1, 015018 (2019)
%doi:10.1103/PhysRevD.99.015018
[arXiv:1809.01876 [hep-ph]].
%110 citations counted in INSPIRE as of 19 Aug 2022

%\cite{LHCb:2015nkv}
\bibitem{LHCb:2015nkv}
R.~Aaij \textit{et al.} [LHCb],
%``Search for hidden-sector bosons in $B^0 \!\to K^{*0}\mu^+\mu^-$ decays,''
Phys. Rev. Lett. \textbf{115}, no.16, 161802 (2015)
% doi:10.1103/PhysRevLett.115.161802
[arXiv:1508.04094 [hep-ex]].
%170 citations counted in INSPIRE as of 23 Aug 2022

%\cite{LHCb:2016awg}
\bibitem{LHCb:2016awg}
R.~Aaij \textit{et al.} [LHCb],
%``Search for long-lived scalar particles in $B^+ \to K^+ \chi (\mu^+\mu^-)$ decays,''
Phys. Rev. D \textbf{95}, no.7, 071101 (2017)
% doi:10.1103/PhysRevD.95.071101
[arXiv:1612.07818 [hep-ex]].
%132 citations counted in INSPIRE as of 23 Aug 2022
  
%\cite{BNL-E949:2009dza}
\bibitem{BNL-E949:2009dza}
A.~V.~Artamonov \textit{et al.} [BNL-E949],
%``Study of the decay $K^+\to\pi^+\nu \bar\nu$ in the momentum region $140 < P_\pi < 199$ MeV/c,''
Phys. Rev. D \textbf{79}, 092004 (2009)
% doi:10.1103/PhysRevD.79.092004
[arXiv:0903.0030 [hep-ex]].
%430 citations counted in INSPIRE as of 24 Aug 2022

%\cite{Gorbunov:2021ccu}
\bibitem{Gorbunov:2021ccu}
D.~Gorbunov, I.~Krasnov and S.~Suvorov,
%``Constraints on light scalars from PS191 results,''
Phys. Lett. B \textbf{820}, 136524 (2021)
% doi:10.1016/j.physletb.2021.136524
[arXiv:2105.11102 [hep-ph]].
%16 citations counted in INSPIRE as of 11 Oct 2022

%\cite{Bondarenko:2019vrb}
\bibitem{Bondarenko:2019vrb}
K.~Bondarenko, A.~Boyarsky, T.~Bringmann, M.~Hufnagel, K.~Schmidt-Hoberg and A.~Sokolenko,
%``Direct detection and complementary constraints for sub-GeV dark matter,''
JHEP \textbf{03}, 118 (2020)
% doi:10.1007/JHEP03(2020)118
[arXiv:1909.08632 [hep-ph]].
%64 citations counted in INSPIRE as of 25 Aug 2022

%\cite{CHARM:1985anb}
\bibitem{CHARM:1985anb}
F.~Bergsma \textit{et al.} [CHARM],
%``Search for Axion Like Particle Production in 400-{GeV} Proton - Copper Interactions,''
Phys. Lett. B \textbf{157}, 458-462 (1985)
% doi:10.1016/0370-2693(85)90400-9
%259 citations counted in INSPIRE as of 25 Aug 2022


%\cite{Billard:2013qya}
\bibitem{Billard:2013qya}
J.~Billard, L.~Strigari and E.~Figueroa-Feliciano,
%``Implication of neutrino backgrounds on the reach of next generation dark matter direct detection experiments,''
Phys. Rev. D \textbf{89}, no.2, 023524 (2014)
% doi:10.1103/PhysRevD.89.023524
[arXiv:1307.5458 [hep-ph]].
%630 citations counted in INSPIRE as of 23 Aug 2022


%\cite{Kawasaki:2000en}
\bibitem{Kawasaki:2000en}
M.~Kawasaki, K.~Kohri and N.~Sugiyama,
%``MeV scale reheating temperature and thermalization of neutrino background,''
Phys. Rev. D \textbf{62}, 023506 (2000)
% doi:10.1103/PhysRevD.62.023506
[arXiv:astro-ph/0002127 [astro-ph]].
%393 citations counted in INSPIRE as of 14 Sep 2022


%\cite{Buschmann:2014sia}
\bibitem{Buschmann:2014sia}
M.~Buschmann, D.~Goncalves, S.~Kuttimalai, M.~Schonherr, F.~Krauss and T.~Plehn,
%``Mass Effects in the Higgs-Gluon Coupling: Boosted vs Off-Shell Production,''
JHEP \textbf{02}, 038 (2015)
% doi:10.1007/JHEP02(2015)038
[arXiv:1410.5806 [hep-ph]].
%132 citations counted in INSPIRE as of 18 Aug 2022


%\cite{Djouadi:2005gi}
\bibitem{Djouadi:2005gi} 
  A.~Djouadi,
%  ``The Anatomy of electro-weak symmetry breaking. I: The Higgs boson in the standard model,''
  Phys.\ Rept.\  {\bf 457}, 1 (2008)
 % doi:10.1016/j.physrep.2007.10.004
  [hep-ph/0503172].
  %%CITATION = doi:10.1016/j.physrep.2007.10.004;%%
  %1341 citations counted in INSPIRE as of 15 Feb 2019
 
 %\cite{Djouadi:2005gj}
\bibitem{Djouadi:2005gj}
A.~Djouadi,
%``The Anatomy of electro-weak symmetry breaking. II. The Higgs bosons in the minimal supersymmetric model,''
Phys. Rept. \textbf{459}, 1-241 (2008)
%doi:10.1016/j.physrep.2007.10.005
[arXiv:hep-ph/0503173 [hep-ph]].
%1508 citations counted in INSPIRE as of 18 Aug 2022 

%\cite{Spira:1995rr}
\bibitem{Spira:1995rr}
M.~Spira, A.~Djouadi, D.~Graudenz and P.~M.~Zerwas,
%``Higgs boson production at the LHC,''
Nucl. Phys. B \textbf{453}, 17-82 (1995)
% doi:10.1016/0550-3213(95)00379-7
[arXiv:hep-ph/9504378 [hep-ph]].
%1524 citations counted in INSPIRE as of 19 Aug 2022

%\cite{Bringmann:2006mu}
\bibitem{Bringmann:2006mu} 
  T.~Bringmann and S.~Hofmann,
 % ``Thermal decoupling of WIMPs from first principles,''
  JCAP {\bf 0704}, 016 (2007)
  Erratum: [JCAP {\bf 1603}, no. 03, E02 (2016)]
%  doi:10.1088/1475-7516/2007/04/016, 10.1088/1475-7516/2016/03/E02
  [hep-ph/0612238].
  %%CITATION = doi:10.1088/1475-7516/2007/04/016, 10.1088/1475-7516/2016/03/E02;%%
  %115 citations counted in INSPIRE as of 05 Mar 2019
 
 %\cite{Bringmann:2009vf}
\bibitem{Bringmann:2009vf} 
  T.~Bringmann,
%  ``Particle Models and the Small-Scale Structure of Dark Matter,''
  New J.\ Phys.\  {\bf 11}, 105027 (2009)
%  doi:10.1088/1367-2630/11/10/105027
  [arXiv:0903.0189 [astro-ph.CO]].
  %%CITATION = doi:10.1088/1367-2630/11/10/105027;%%
  %188 citations counted in INSPIRE as of 05 Mar 2019
  
%\cite{Gondolo:2012vh}
\bibitem{Gondolo:2012vh} 
  P.~Gondolo, J.~Hisano and K.~Kadota,
%  ``The Effect of quark interactions on dark matter kinetic decoupling and the mass of the smallest dark halos,''
  Phys.\ Rev.\ D {\bf 86}, 083523 (2012)
 % doi:10.1103/PhysRevD.86.083523
  [arXiv:1205.1914 [hep-ph]].
  %%CITATION = doi:10.1103/PhysRevD.86.083523;%%
  %34 citations counted in INSPIRE as of 05 Mar 2019
  
  %\cite{Visinelli:2015eka}
\bibitem{Visinelli:2015eka} 
  L.~Visinelli and P.~Gondolo,
%  ``Kinetic decoupling of WIMPs: analytic expressions,''
  Phys.\ Rev.\ D {\bf 91}, no. 8, 083526 (2015)
%  doi:10.1103/PhysRevD.91.083526
  [arXiv:1501.02233 [astro-ph.CO]].
  %%CITATION = doi:10.1103/PhysRevD.91.083526;%%
  %15 citations counted in INSPIRE as of 05 Mar 2019
  
%\cite{Binder:2016pnr}
\bibitem{Binder:2016pnr} 
  T.~Binder, L.~Covi, A.~Kamada, H.~Murayama, T.~Takahashi and N.~Yoshida,
 % ``Matter Power Spectrum in Hidden Neutrino Interacting Dark Matter Models: A Closer Look at the Collision Term,''
  JCAP {\bf 1611}, 043 (2016)
%  doi:10.1088/1475-7516/2016/11/043
  [arXiv:1602.07624 [hep-ph]].
  %%CITATION = doi:10.1088/1475-7516/2016/11/043;%%
  %32 citations counted in INSPIRE as of 06 Mar 2019

%\cite{Binder:2017rgn}
\bibitem{Binder:2017rgn} 
  T.~Binder, T.~Bringmann, M.~Gustafsson and A.~Hryczuk,
%  ``Early kinetic decoupling of dark matter: when the standard way of calculating the thermal relic density fails,''
  Phys.\ Rev.\ D {\bf 96}, no. 11, 115010 (2017)
%  doi:10.1103/PhysRevD.96.115010
  [arXiv:1706.07433 [astro-ph.CO]].
  %%CITATION = doi:10.1103/PhysRevD.96.115010;%%
  %23 citations counted in INSPIRE as of 21 Mar 2019



\end{thebibliography}
\end{document}